# Size, Shape, and Material matter: All-optical Mie void sensor for complex nanoplastic mixtures

D. Ludescher [1,*], J. Schwab [1], S. Arslan [1], E. Kubacki [2], M. Ubl [1], M. Retsch [2,3], H. Giessen [1], and M. Hentschel [1,*]

[1] *4th Physics Institute and Research Center SCoPE, University of Stuttgart, 70569 Stuttgart, Germany*
[2] *Department of Chemistry, Physical Chemistry I, University of Bayreuth, 95447 Bayreuth, Germany*
[3] *Bavarian Polymer Institute, Bayreuth Center for Colloids and Interfaces, Bayreuther Institut für Makromolekülforschung, and Bavarian Center for Battery Technology (BayBatt), University of Bayreuth, 95447 Bayreuth, Germany*
[*] *Corresponding authors: Mario Hentschel (m.hentschel@pi4.uni-stuttgart.de) and Dominik Ludescher (d.ludescher@pi4.uni-stuttgart.de)*

## Abstract

The fragmentation of plastic debris and the direct release of nanoplastics have emerged as a pressing ecological concern. Once dispersed, these enduring particles infiltrate food webs, accumulate within organisms, and bind toxic co-contaminants, posing long-term risks to ecosystems and human health. Despite growing awareness, the detection and characterization of nanoplastics remain highly challenging due to their minute size. Moreover, obtaining additional critical information, such as the particle shape or material composition, further exacerbates these detection hurdles. Conventional analytical techniques capable of providing more detailed information often demand substantial experimental and lab-bound effort, costly instrumentation, and lengthy measurement times.
Here, we introduce a novel photonic sensing platform based on nanoscale voids that enables the simultaneous material- and morphology-sensitive detection of particles below 500 nm. Arrays of voids embedded in a high-refractive-index material act in parallel as both sorting elements and direct color reporters. Spherical and elongated particles are selectively trapped in circular and elliptical voids, while polymer types such as PS, PMMA, and PET are distinguished via the specific color signatures arising from their refractive index contrasts. This approach offers a cheap and scalable route toward rapid optical identification of nanoplastics in complex environmental and biological settings. Its compatibility with quick, high-throughput analysis positions it as a promising tool for real-time monitoring and comparative studies of heterogeneous nanoplastic populations.

Scenes of plastic pollution are omnipresent around us: once pristine beaches are rendered unrecognizable by accumulating plastic trash; marine animals such as turtles and fish are entangled in discarded fishing nets or plastic bags and bottles; floating islands of debris are drifting across the world's oceans. It is now unequivocal that plastic pollution constitutes one of the most severe environmental challenges of the twenty-first century[1–5]. While plastics owe their widespread use to desirable material properties such as low weight, high durability, and cost-effective manufacturing, this resulting pollution demands urgent intervention[6–8]. The global distribution of plastic debris, spanning from the highest remote mountain regions[9,10] to the deepest ocean trenches[11–14], underscores both the scale and persistence of this issue.

Although these images of extensive macro- and microplastic pollution vividly illustrate the trouble of managing plastic waste, scientists agree that an even more severe threat is emerging, one that is invisible to the naked eye, namely nanoplastics[15–17]. These particles, defined as being smaller than 1 μm, are considered particularly hazardous due to their enhanced mobility, increased bioavailability, and high propensity for cellular uptake[18]. Nanoplastics originate either from primary sources, such as their intentional use as additives in consumer products, or more commonly from the progressive fragmentation of larger plastic debris in the environment. They are now ubiquitous in freshwater systems[19,20], the food chain[21–23], and the atmosphere[24,25], resulting in continuous and largely unavoidable human exposure. Once internalized, nanoplastics can bioaccumulate in organs, traverse biological barriers such as the blood-brain barrier, and, more critically, act as carriers for highly toxic contaminants that may subsequently be released into the human body[26–30].

Consequently, there is an urgent need for new, reliable detection methods. In the last few years, a pivotal shift has occurred from laboratory-bound analytical systems toward field-deployable devices that are cost-effective to fabricate and straightforward to operate[31–34]. However, owing to their extremely small size, the detection of nanoplastics continues to face substantial analytical challenges. To overcome limitations such as low throughput and insufficient spatial resolution, complementary detection methods, including optical inspection, spectroscopic techniques, and thermal analysis, are routinely integrated to leverage their respective strengths and achieve robust, reliable nanoplastic detection. Microscopy-based techniques, such as optical and scanning electron microscopy (SEM), are well-suited for the analysis of larger particles and enable single-particle investigation. Nevertheless, they are often constrained by limited capacity and typically provide limited information on material composition[31,35,36]. In contrast, Fourier-transform infrared and Raman spectroscopy are widely regarded as the gold standard for polymer identification, as they probe molecular fingerprints that allow direct determination of polymer type[37,38]. However, these spectroscopic techniques suffer from limited spatial resolution, require expert interpretation, and are generally restricted to laboratory settings[32,39–42]. On the other hand, thermal analytical approaches, such as pyrolysis gas chromatography-mass spectrometry, offer rapid and highly sensitive detection, but rely on complex sample preparation and face challenges in scalability[43–45]. Taken together, these constraints make the combination of different techniques inevitable and emphasize the need for the development of novel concepts that expand and diversify the existing analytical toolbox for nanoplastic detection.

Very recently, a novel detection strategy based on Mie-type void resonators embedded in high-refractive index substrates such as gallium arsenide or silicon has been introduced[46,47]. These resonators consist of open, cylindrical cavities etched into the substrate surface, where light is confined at discrete wavelengths, determined by the cavity geometry and the refractive index of the medium filling the void[48]. This strong dependence enables the detection of particles trapped inside the cavities and allows their sorting, sizing, and counting based on the particle properties. While this approach benefits from its simplicity, requiring only a standard optical microscope, further research is needed before low-cost, field-deployable operation by non-specialized personnel becomes feasible. To date, investigations have been limited to a highly simplified model system consisting of only spherical polystyrene particles, which restricts the reliable assessment of more complex sample matrices[49,50].

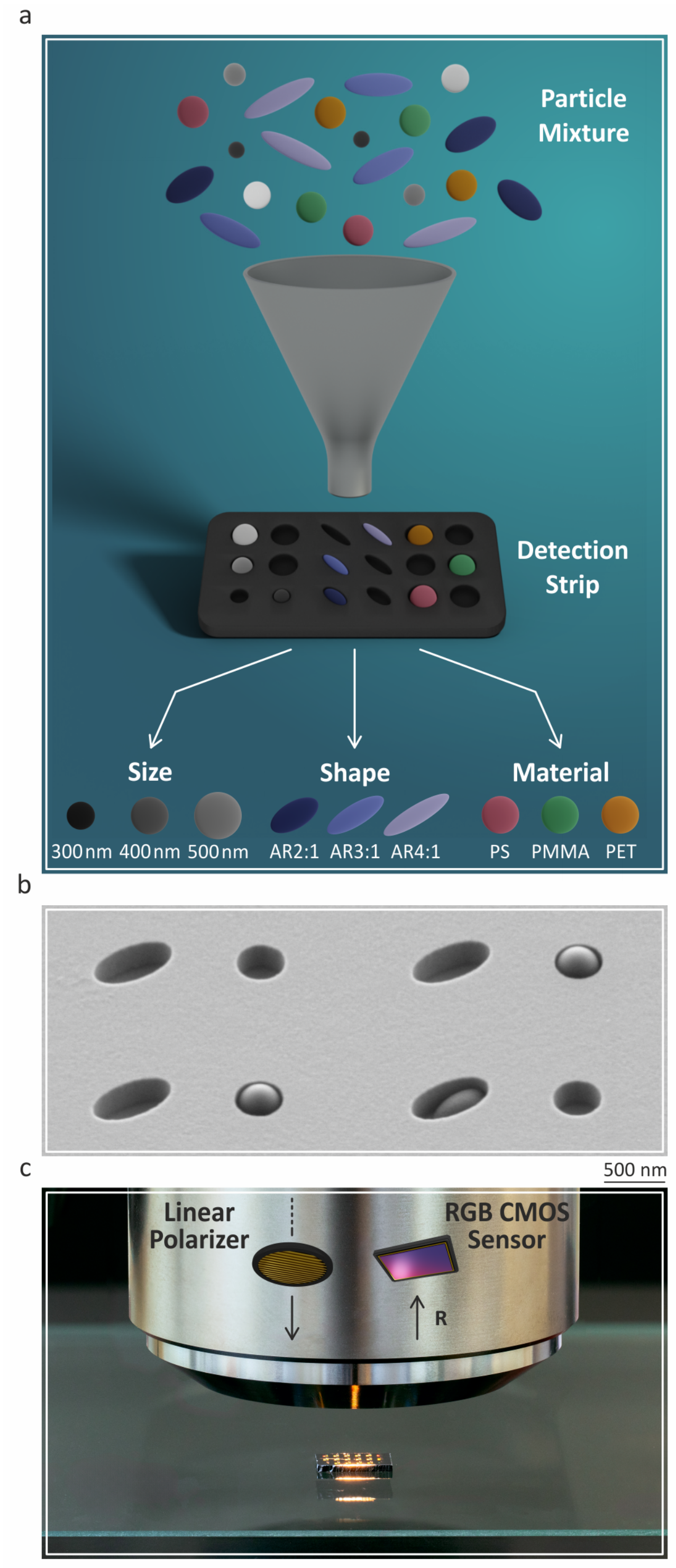

**Figure 1: Sensor principle.** **a**, Schematic illustration of the measurement principle. A mixture of spherical particles of different sizes or different materials, as well as ellipsoidal particles with varying aspect ratios (ARs), is analyzed. The detection strip, consisting of void arrays, is fabricated in gallium arsenide substrates and enables the material- and shape-sensitive particle detection. **b**, SEM image of voids presenting trapped round and ellipsoidal particles. **c**, Photograph of the measurement. A linear polarizer combined with a standard microscope objective is used to image the detection strips. The linear polarizer (left) and the RGB CMOS sensor (right) are indicated schematically.

Generally, nanoplastics can be classified according to their size, shape, material type, and composition. Here, we demonstrate simultaneous sorting of particles by both size and shape, enabled by the intrinsic sorting functionality of the Mie void resonators themselves[35]. Furthermore, by exploiting refractive-index contrasts, we discriminate between visually indistinguishable particles through their characteristic resonance colors. This enables composition-sensitive detection using only an optical microscope combined with a simple RGB color analysis. Finally, the detection fidelity can be significantly enhanced by implementing advanced and automated image analysis, incorporating polarization-resolved imaging.

The ultimate goal of this work is the realization of a nanoplastic test strip compatible with routine use, enabling purely optical extraction of particle size, size distribution, material type, and composition from environmentally derived samples using standard optical microscopy. To this end, we extend the functionality of Mie void-based nanoplastic detection by demonstrating the simultaneous identification of particles composed of different materials (PS, PMMA, and PET), as well as the differentiation between spherical and ellipsoidal particles with varying aspect ratios.

## Results

### General detection mechanism

In this work, we employ a detection strip composed of arrays of dielectric void resonators embedded in a high-refractive-index medium, which simultaneously function as size- and shape-selective trapping sites and as optical reporters. Upon deposition of a liquid sample containing sub-micrometre nanoplastics, particles are preferentially captured in voids whose geometry closely matches their size and shape, thereby maximizing van der Waals interactions. When the void dimensions and particle sizes deviate from each other, the interaction forces are reduced, increasing the likelihood of particle removal during subsequent cleaning steps (see Methods). Previous experiments, however, have shown that for particles with a nominal diameter of 350 nm, a diameter mismatch of approximately $\pm$80 nm can still result in trapping efficiencies of around 5 - 10 %[49]. To maximize trapping efficiencies, the void dimensions are therefore designed to closely match the particle size. Each detection site supports a localized Mie void resonance that is highly sensitive to changes in the effective refractive index within the cavity[51,52]. The capture of particles with different sizes or material properties therefore induces a deterministic spectral shift of the resonance, which manifests as a distinct color change observable using an optical microscope. This approach enables size- and shape-selective analysis combined with a material-sensitive optical readout. Please note that the voids used in the experiments were not optimized for specific optical resonances but were instead designed to match the expected particle dimensions in the specimen. The usage of broadband white-light microscopy then ensures the presence of optical resonances.

Figure 1a schematically illustrates the detection principle. A heterogeneous mixture of plastic particles is analysed using the detection strip, allowing particles with different sizes, shapes, and material compositions to be probed simultaneously on the same platform.

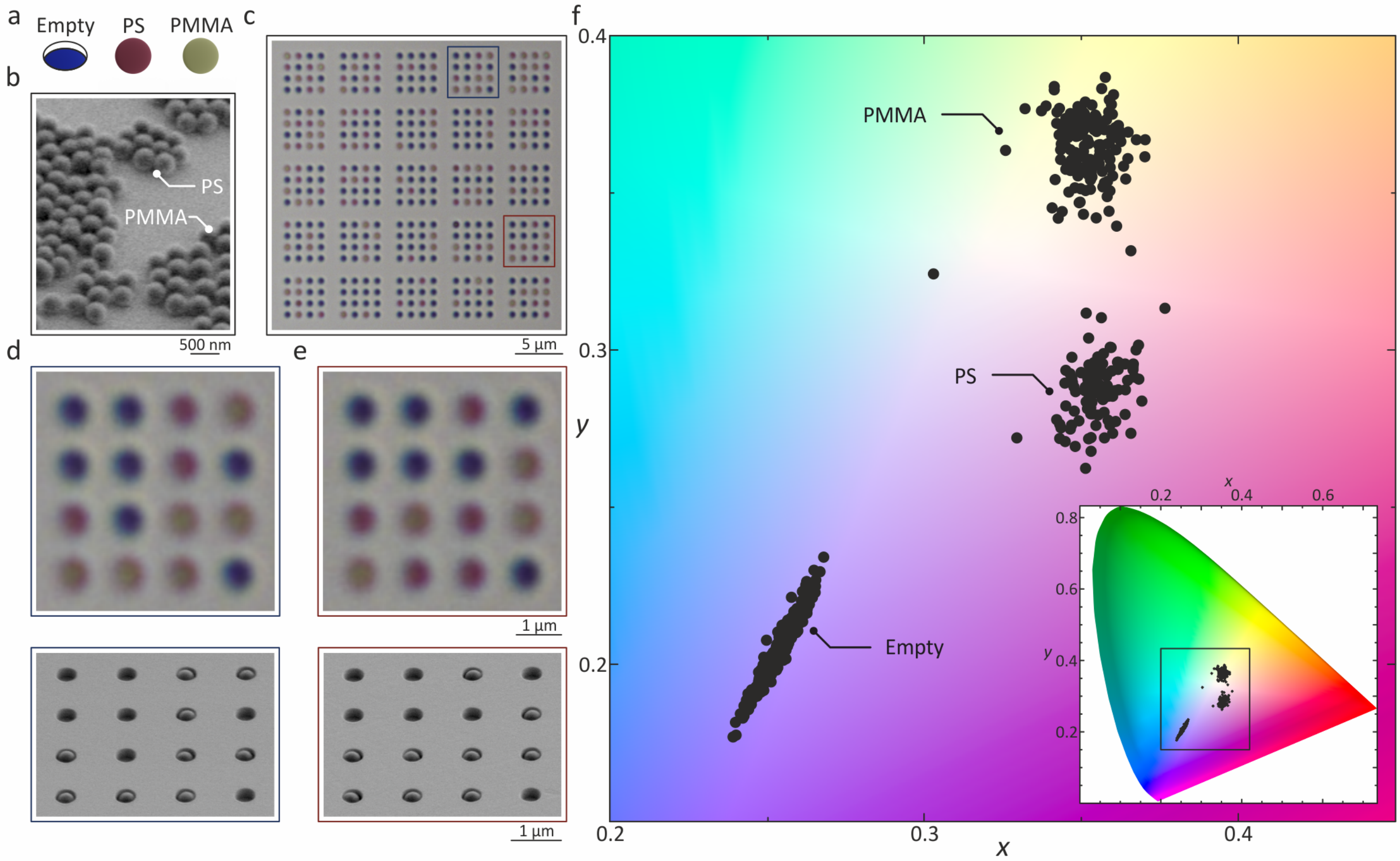

**Figure 2: Distinguishing different nanoplastic materials. a**, Different polymer nanoparticles (PS and PMMA) with similar diameters of about 300 nm can be distinguished by their characteristic Mie void colors. **b**, SEM image of the particle mixture deposited onto the Mie void sample. **c**, Optical microscope image of the void array, consisting of 5 x 5 fields, each containing 16 voids. Empty voids remain blue, while the filled voids become red (PS) or orange (PMMA). **d & e**, Magnified microscope and SEM views of two distinct fields, highlighting the material-dependent color contrast. **f**, Color representation of the detected void colors across the entire detection strip. (Inset) Full CIE-1931 color diagram.

The SEM image in Figure 1b depicts the principle of shape-selective particle discrimination. As demonstrated in later experiments, spherical and ellipsoidal particles are preferentially trapped only in voids with matching geometries, while particles in mismatched voids are readily removed during the cleaning process using ultrasonication and water rinsing. Figure 1c presents a photograph of the experimental setup, consisting of a standard optical microscope equipped with a rotatable linear polarizer and a conventional CMOS camera. This configuration enables the detection of color shifts associated with empty and particle-filled voids and forms the basis for the subsequent RGB-based colorimetric analysis.

## Material-sensitive detection

The SEM images depicted in Supplementary Fig. 1 demonstrate that polystyrene (PS) and poly(methyl methacrylate) (PMMA) particles cannot be reliably differentiated based on morphology alone. In particular, Supplementary Fig. 1c highlights that visual differences between the two materials are minimal, rendering SEM imaging insufficient for unambiguous material identification. Thus, to demonstrate the capability of material-sensitive detection of our plat-form, Figure 2 presents voids filled with PS or PMMA particles of similar nominal diameter of 300 nm. They can be clearly discriminated solely using their optical color response. Spherical PS (n = 1.59 @ 633 nm) and PMMA particles (n = 1.49 @ 633 nm)[53], which predominantly differ in their refractive index, were deposited onto the detection strip. To further characterize the particle ensembles, dynamic light scattering (DLS) measurements were performed (Supplementary Fig. 2), yielding average diameters of 363 nm for PS and 374 nm for PMMA, with a polydispersity index (PDI) of 0.099 and 0.005 , respectively. These determined size variations for particles with nominal diameters of 300 nm indicate that intrinsic differences in the particle dimensions dominate over other sources of uncertainty, such as fabrication-induced inaccuracies of the sensor structures.

Figure 2a schematically summarizes the three relevant cases of empty voids (blue), PS-filled voids (red), and PMMA-filled voids (yellow), while Figure 2b depicts an SEM image of the deposited sample prior to the cleaning step, once more indicating the incapability of visually distinguishing particles of different material types. A section of the full detection strip used for the experiment is presented in Figure 2c. In Figure 2d, e, two representative detection fields, each comprising $4 \times 4$ voids, together with the corresponding SEM images, are displayed. To unambiguously correlate the observed colors with the corresponding materials, reference measurements using exclusively PS (Supplementary Fig. 3) or PMMA particles (Supplementary Fig. 4) were performed. The voids utilized in this experiment have a diameter of approximately 410 nm, a depth of 310 nm, and a center-to-center distance of 1.2 µm.

The resulting colorimetric data are summarized in the CIE 1931 color space in Figure 2f, with the full diagram revealed in the inset. In both representations, three well-separated clusters are observed, corresponding to the three named regimes, thereby demonstrating clear material-specific discrimination based on the optical response of the detection strip. Overall, more than 600 voids have been analyzed in their color appearance. The color readout is performed via an automated code that identifies the periodic voids and determines the average color of the voids (see Methods).

In the CIE chromaticity diagram, a single outlier is observed. This data point appears near the center of the diagram, in close proximity to the light-blue region, and cannot be assigned to any of the three distinct clusters. Inspection of the corresponding SEM and optical microscopy images (Supplementary Fig. 5) reveals that this deviation originates from a dirt particle partially occupying the void. The presence of this impurity perturbs the local optical resonance, resulting in an observable shift of the perceived color. Such incorrectly filled voids can therefore be readily distinguished from empty or correctly filled voids solely based on their resonance color following initial verification by SEM imaging. Moreover, optical control experiments conducted on flat GaAs substrates without voids are presented in Supplementary Figs. 6 and 7. Here, PS particles with different nominal diameters (300 nm and 500 nm) and a mixture of PS and PMMA particles with nominal diameters of 300 nm are deposited. Analysis of the optical microscope images reveals that the refractive-index-dependent optical response of the Mie voids is crucial for enabling precise nanoplastic detection.

## Shape-sensitive detection

After establishing the material sensitivity of the platform, we next highlight its intrinsic shape sensitivity. At this point, we simultaneously investigate spherical and ellipsoidal particles. Ellipsoidal particles are fabricated by mechanically elongating spherical PS particles embedded in a PVA matrix (see Methods), yielding an aspect ratio (AR) of about 2:1, defined as the quotient of the long and short axes.

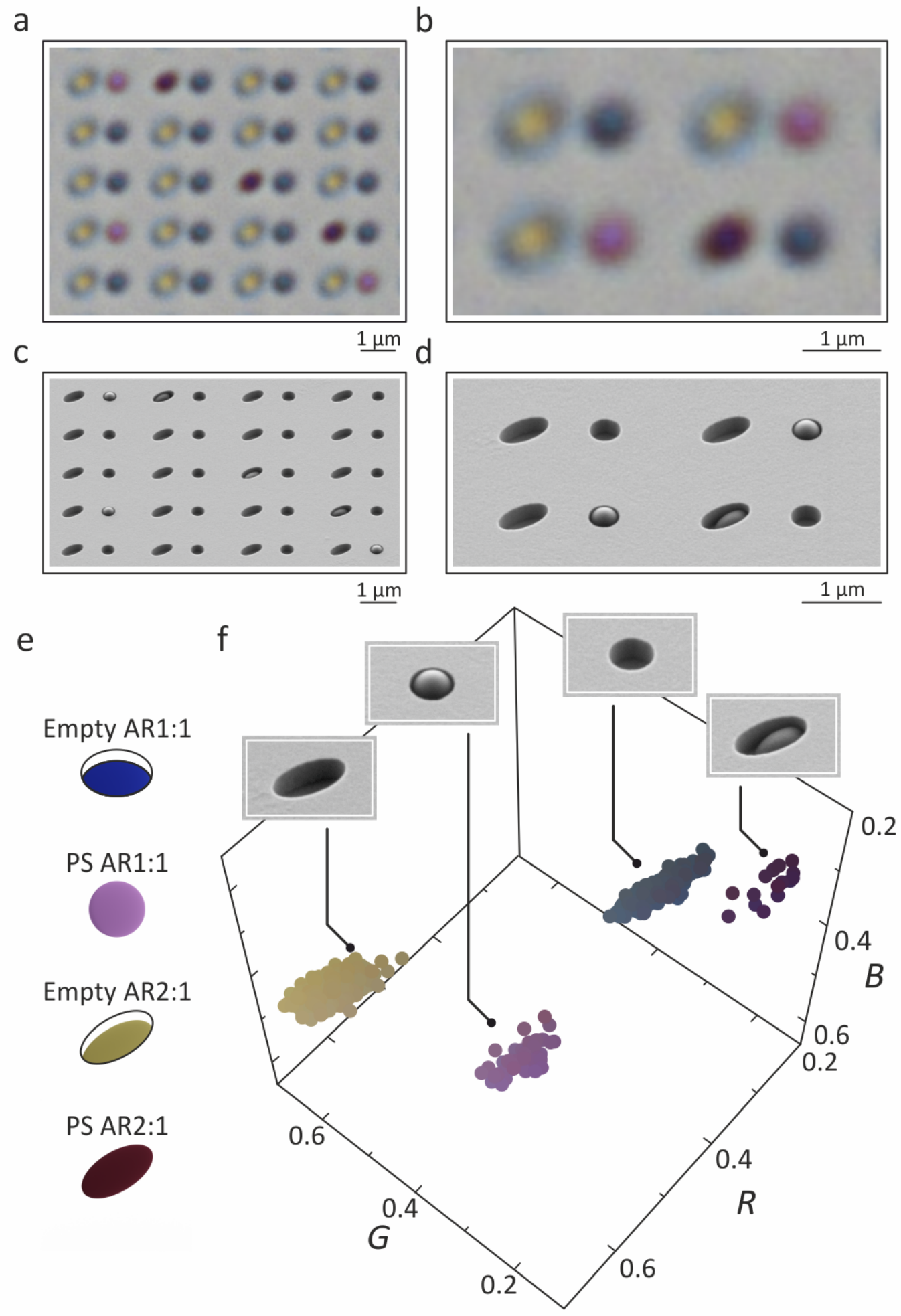

**Figure 3: Shape-selective nanoparticle sorting. a & b**, Optical microscope and **c & d**, SEM images of the detection strip with elliptical (left rows) and round (right rows) voids. Ellipsoidal particles are trapped only in elliptical voids, while spherical particles (350 nm diameter) only occupy round voids. Linear polarization along the short axis of the elliptical voids is utilized for image capturing. **e**, Colors of the empty round and elliptical voids and the deposited mixture of round PS particles and ellipsoidal PS particles with an aspect ratio of 2:1. **f**, RGB color mapping reveals four distinct clusters: empty round (blue), filled round (pink), empty elliptical (yellow), and filled elliptical (dark-red). SEM images around the RGB cube illustrate each void state.

When mixtures of spherical and ellipsoidal particles are deposited onto detection strips containing both round and elongated voids, a pronounced shape selectivity is observed. Spherical particles are exclusively trapped in round voids, whereas ellipsoidal particles preferentially occupy elongated voids. Particles residing in mismatched voids experience reduced adhesion forces and are therefore removed during the ultrasonication-based cleaning step. As a result, the voids act not only as passive trapping sites but also as sorting elements and all-optical reporters of particle shape.

Optical microscope images of the detection strip are presented in Figure 3a and Figure 3b, displaying in periodic arrangement elongated voids (left) and round voids (right), respectively. Here, linear polarization in the illumination path of the optical microscope along the short axis of the voids is used to increase the image contrast. The corresponding SEM images are presented in Figure 3c and Figure 3d. The circular voids possess a diameter of about 400 nm, while the elliptical voids have major and minor axis lengths of approximately 600 nm and 300 nm. The void depth is approximately 250 nm, and the center-to-center distance between adjacent voids is 1 µm. Figure 3e summarizes the four distinct configurations encountered in this experiment: empty and filled round voids (blue and pink), as well as empty and filled elongated voids (yellow and red). All particles used in this study are composed of PS. As before, an optical control experiment on a flat GaAs substrate, displayed in Supplementary Fig. 8, is performed to further clarify the importance of the void structures. Particles with different shapes cannot be reliably distinguished using a bare substrate without fabricated voids.

In the previous figure, the CIE color space was used to visualize the colors associated with empty and filled voids. Although CIE is among the most widely used color representations, the analysis is not limited to this choice. A range of alternative color spaces is available, and an appropriate selection can enhance the separation between distinct color clusters. One such alternative is the three-dimensional RGB color space, in which each axis corresponds to an individual color channel, which is well matched to the detector pixels of the utilized CMOS camera. To illustrate the influence of the chosen color space, identical void colors extracted from Figure 3 are mapped in both RGB and CIE representations displayed in Supplementary Fig. 9. While the RGB color map used in Fig. 3f reveals four clearly separated color clusters, with corresponding SEM images indicating the underlying void states, the CIE diagram is incapable of discriminating the distinct clusters. The absence of intermediate data points between clusters demonstrates that the detection strip enables reliable shape-based sorting of particle mixtures, while simultaneously discriminating empty from filled voids through a purely optical readout.

While spherical particles are not observed to be trapped in elliptical voids in our experiment, ellipsoidal particles were trapped tip-first in circular voids in very rare cases (Supplementary Fig. 10). While such events occur with very low probability and are typically resolved during the cleaning process, occasional residual particles remain. In total, 7500 circular voids were analyzed, among which 11 contained an ellipsoidal particle trapped tip-first, corresponding to a mis-trapping probability of 0.15 %. Importantly, even in these cases, the effective refractive index perturbation differs from that induced by a spherical particle due to the altered modal overlap. Consequently, the resulting optical response, and thus the observed color, remains distinct, enabling unambiguous differentiation between spherical and ellipsoidal particles in circular voids after initial reference measurements.

To further increase the level of complexity beyond shape discrimination between spherical and ellipsoidal particles, we investigate ellipsoidal particles with different aspect ratios. Specifically, particles with ARs of 2:1, 3:1, and 4:1 are studied. All three particle populations are deposited simultaneously onto a single detection strip containing elliptical voids with dimensions and ARs matched to those of the particles. Detailed statistics of the particle dimensions and AR-distributions are provided in Supplementary Fig. 11. This analysis yields experimental ARs of $2.02(\pm 0.16){:}1$, $2.85(\pm 0.20){:}1$, and $3.74(\pm 0.23){:}1$.

Figure 4a summarizes the four configurations observed on the detection strip: empty voids (yellow) and voids filled with ellipsoidal particles, where increasing ARs are visible by progressively darker red colors. An SEM image of the sample prior to the cleaning step is presented in Figure 4b, where particles of all three ARs are present simultaneously on the surface. Figures 4c-e present the individual investigations for AR = 2:1 (top), 3:1 (middle), and 4:1 (bottom), respectively. For each AR, an overview SEM image (left) is displayed alongside a corresponding optical microscope image (center) and an SEM image of the same region

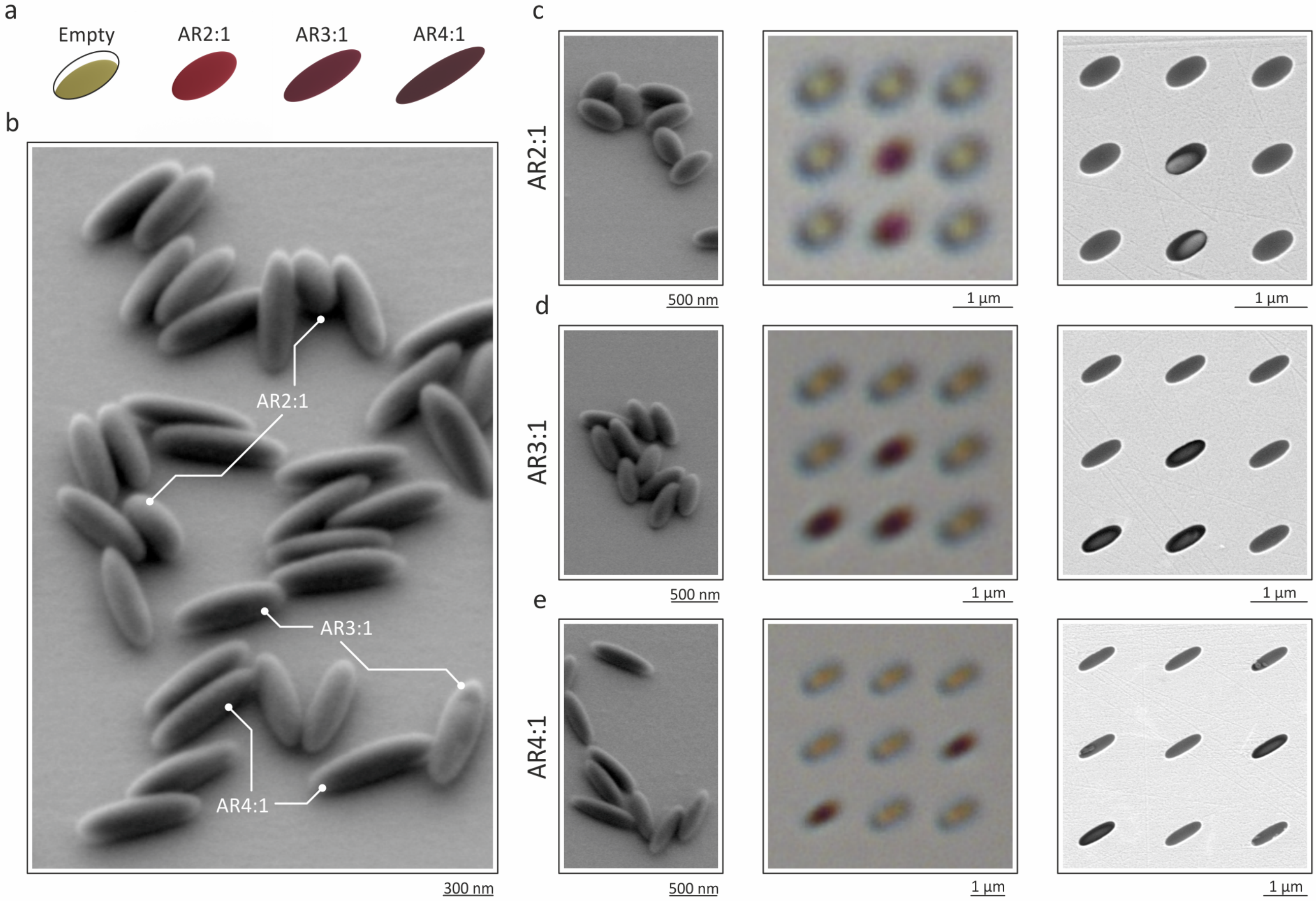

**Figure 4: Ellipsoidal nanoparticles with different aspect ratios.** **a**, Deposited mixture of ellipsoidal particles with three different aspect ratios. **b**, SEM image of the utilized mixture. **c-e**, Ellipsoidal particles with aspect ratios of 2:1, 3:1, and 4:1 are used: (left) SEM image of the individual particle types, (middle) optical microscope images under linear polarization along the short axis of the void, and (right) SEM images depicting the filled and empty voids.

(right). The optical images were again acquired using linearly polarized illumination oriented along the short axis of the voids. The voids designed for particles with an AR of 2:1 have major and minor axis lengths of 620 nm and 340 nm, respectively. For AR = 3:1, the major and minor axis lengths are 800 nm and 310 nm, respectively. For an AR of 4:1, the corresponding axis lengths are 960 nm and 290 nm. The depth of the voids is approximately 320 nm, and the center-to-center distance increases from 1.2 µm to 1.5 µm and 2.0 µm. In all three cases, filled voids exhibit a pronounced red coloration, whereas empty voids appear yellowish, allowing for straightforward visual discrimination. Importantly, particles with incorrect ARs are not retained in the voids and are efficiently removed during the cleaning process. This demonstrates that the detection strip enables selective trapping and sorting even among particles of identical material composition but different shapes.
As in the previous experiment, rare instances occur in which a particle with a mismatched AR remains trapped after cleaning. Although the adhesion forces in such configurations are typically weak, they can occasionally be sufficient to withstand ultrasonication. An example of this scenario is indicated in Supplementary Fig. 12, where a single ellipsoidal particle with an AR of approximately 3:1 remains trapped in a void designed for AR = 2:1 (upper right 3 $\times$ 3 array). Crucially, even in this case, the effective refractive index perturbation, and thus the optical response, differs from that of correctly matched particles. This results in a lighter red coloration for the mismatched particle, compared to the darker red associated with the correctly trapped particles, enabling unambiguous optical identification.

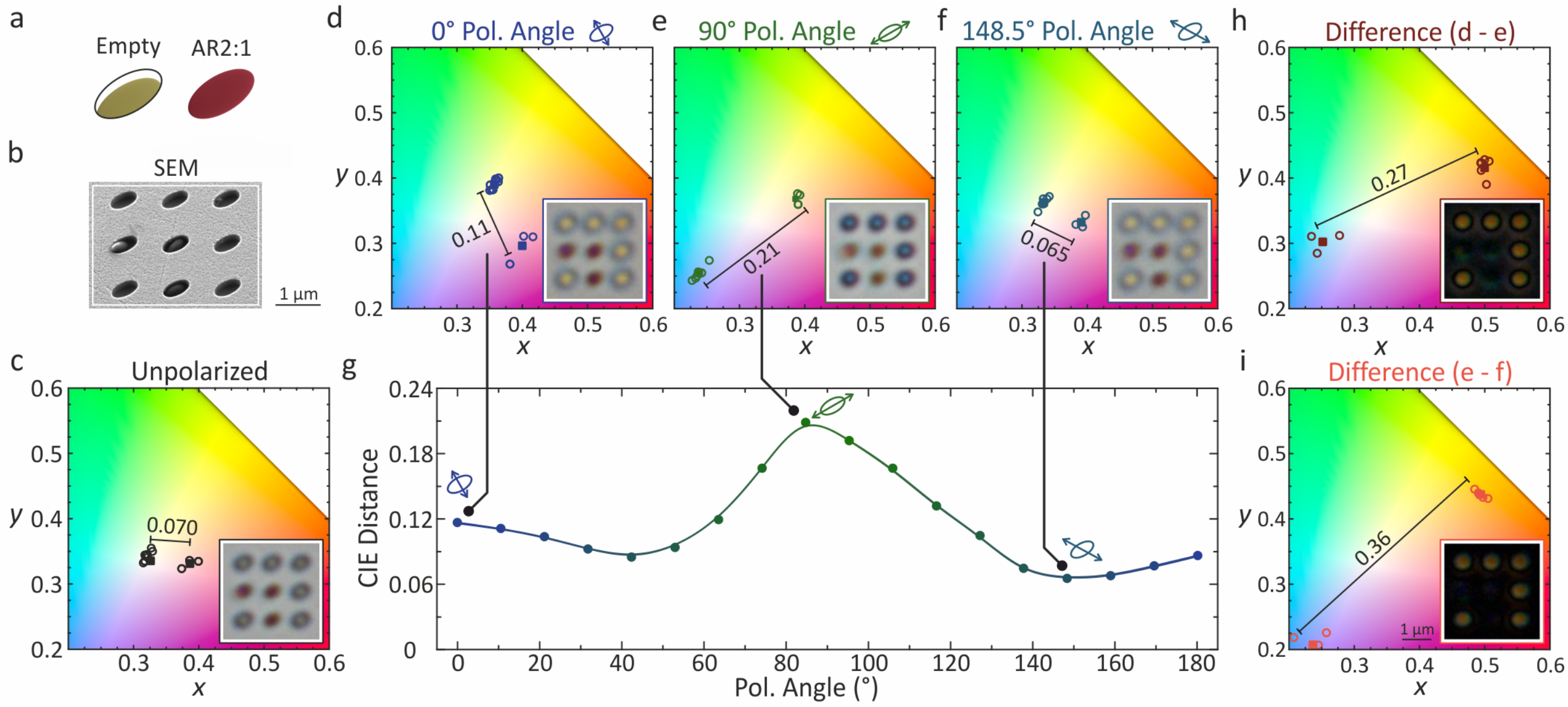

**Figure 5: Polarization-dependent color contrast enhancement.** Detection of ellipsoidal particles is optimized through polarization-sensitive imaging and differential analysis of different polarization states. **a**, The deposited ellipsoidal particles have an aspect ratio of 2:1. **b**, SEM image of the examined detection field containing three filled voids. **c-f**, Four different polarization states are examined. In the CIE-1931 diagrams, represented by the *x* and *y* values, the color of the six empty and the three filled voids is displayed (circles). The average color values (squares) reveal that the color separation depends on the polarization angle. The reported values correspond to the distance between the mean color coordinates. **g**, CIE distance as a function of polarizer angle, showing a maximum color separation for polarization aligned with the long axis of the void and a minimum separation at a polarization angle of 148.5 °. **h & i**, Differential images obtained from images **d** and **e**, and from images **e** and **f**, respectively. The increased distance values demonstrate that differential imaging can enhance the color contrast between empty and filled voids. (Insets) Optical microscope images used for the analysis.

### Advanced and polarization-dependent image analysis

As demonstrated above, the detection strip can be employed as an effective sorting element for particles with different shapes. Moreover, the ellipsoidal particles exhibit polarization-dependent behavior, enabling advanced optical analysis. In certain configurations, the intrinsic color contrast between particle states becomes weak, necessitating additional contrast en-hancement. To address this limitation, we employ polarization-resolved imaging, which enables an amplification of subtle color differences by comparing different polarization states.

Figure 5 illustrates this approach for ellipsoidal particles with an AR of 2:1 deposited on the detection strip, resulting in two distinct configurations: empty and particle-filled voids (Figure 5a). A SEM image of the investigated region is shown in Figure 5b, confirming the presence of six empty and three filled voids. The voids designed for particles with an AR of 2:1 possess a major axis length of approximately 610 nm, a minor axis length of 300 nm, a depth of about 350 nm, and a center-to-center distance of 1.2 µm.
We initially compare four different imaging modalities: unpolarized illumination (Figure 5c), linear polarization aligned along the short axis (Figure 5d), along the long axis of the ellipses (Figure 5e), and with a polarization angle of 148.5 ° (Figure 5f). In addition, the graph in Figure 5g displays the determined CIE distance as a function of the polarization angle, illustrating the sensitivity of the CIE color distance to changes in polarization. The corresponding images used to calculate the polarization-angle-dependent CIE distances are provided in Supplementary Fig. 13. Representing the void colors in the two-dimensional CIE color space enables quantitative comparison through distances between color values. We define the color contrast as the Euclidean distance between two points in the CIE diagram.
Moreover, we calculate the color difference between the two perpendicular polarization states depicted in Figures 5d and 5e, as presented in Figure 5h. We further determine the color difference between the two polarization states shown in Figures 5e and 5f, as depicted in Figure 5i. For each case, the corresponding microscope image is shown as an inset, and the extracted color information is represented in the CIE chromaticity diagram. Individual measurements of the six empty and three filled voids are presented as open circles, while the respective mean values, used for distance calculations, are indicated by filled squares.
Under unpolarized illumination, the color separation between empty and filled voids is modest, yielding a CIE distance of 0.070 . Introducing linear polarization in the illumination path leads to a pronounced enhancement of the color contrast. For polarization aligned along the short axis, the color distance increases to 0.11 and further rises to 0.21 for alignment along the long axis. The smallest color distance of 0.065 is observed at a polarization angle of 148.5 °. In all polarized cases, filled voids consistently exhibit reddish hues, while the empty voids undergo a pronounced color shift from yellow to blue, depending on the polarization orientation.
A substantially larger contrast enhancement is achieved by computing the difference image between two different linear polarization states (Figures 5h and 5i). This difference operation is performed independently for each color channel by subtracting the lower-brightness value from the higher-brightness value. As a result, regions without polarization-dependent color changes appear black, while regions with strong polarization sensitivity appear bright. In this representation, the filled voids exhibit only a weak color response, resulting in dark tones, whereas the empty voids show a pronounced polarization-dependent color change and therefore appear in brighter orange. Notably, the background is completely black in the difference image, as it exhibits no measurable polarization dependence. Here, the difference image was calculated between the orthogonal polarization states, resulting in a color contrast value of 0.27 (Figure 5h). Although this comparison might initially be expected to yield the largest color distance, an even higher value of 0.36 (Figure 5i) is obtained when calculating the difference between the polarization states corresponding to the highest and lowest CIE distances in Figure 5g. The reason for this is that the color contrast is defined as the distance in the CIE diagram, and colorimetric shifts of the resonances induced by polarization variations cannot be predicted intuitively.
Consequently, this simple analysis significantly amplifies small intrinsic color contrasts and enables a clear discrimination between particle-filled and empty voids. Polarization-dependent difference imaging thus provides an efficient strategy to enhance contrast in optical sorting and detection schemes.
A detailed analysis of the polarization sensitivity of round, empty voids is provided in Supplementary Fig. 14. This analysis implies that round voids exhibit no polarization sensitivity.Moreover, in comparison to Figure 5g, the polarization sensitivity of the CIE distance for ideally filled round voids is presented in Supplementary Fig. 15. Here, the CIE distance does not show significant changes as a function of the polarizer angle.

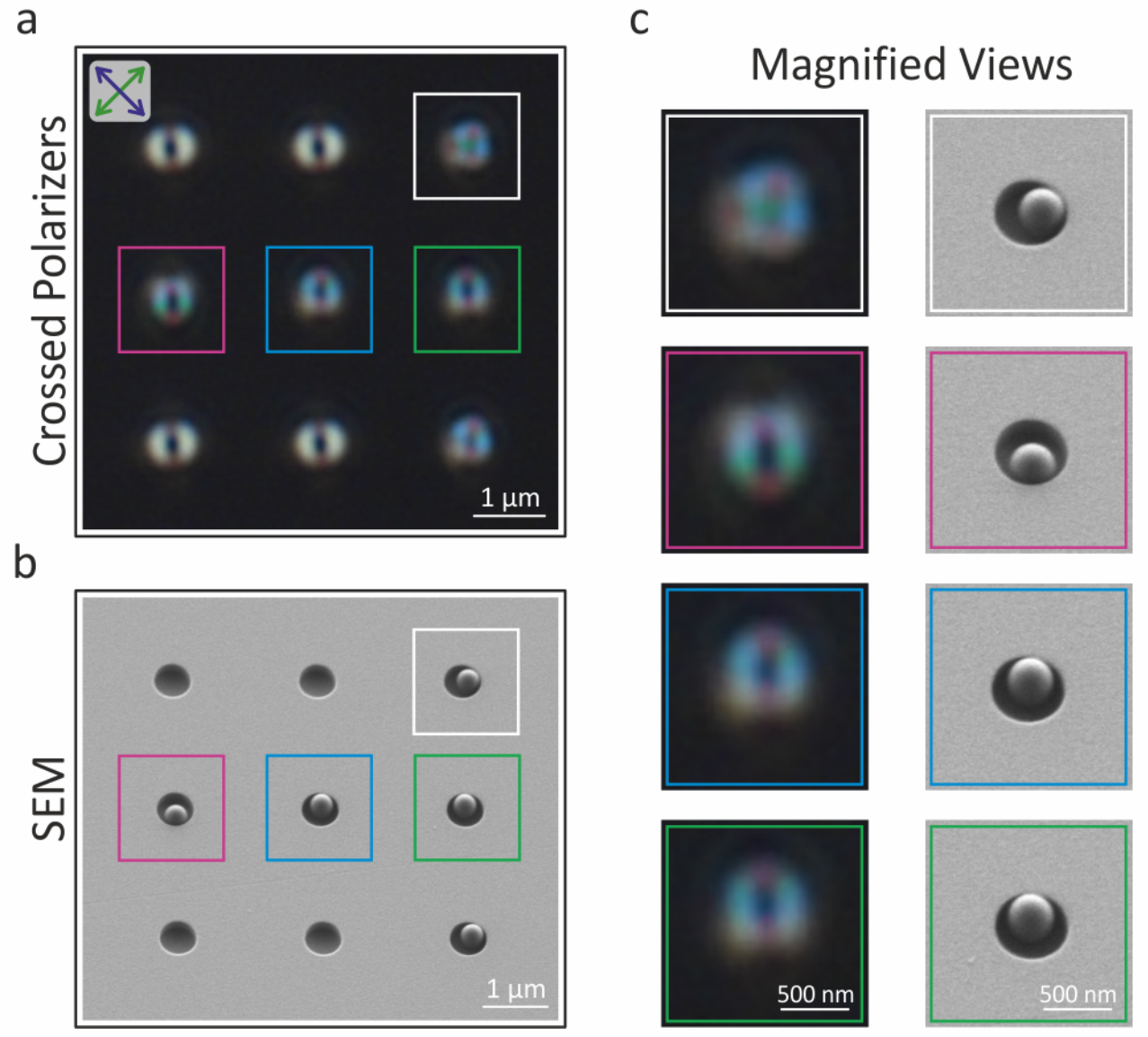

**Figure 6: Crossed-polarizer analysis.** Particle positions within individual voids are analyzed using two crossed linear polarizers. **a**, Optical microscope image acquired under crossed-polarizer illumination. **b**, Corresponding SEM image indicating the different particle positions within the voids. **c**, Magnified views of four distinct voids. The colored frames indicate the corresponding voids in the optical and SEM image. Comparison of the voids marked by the white, pink, and blue frames shows that the optical microscope image already provides information about the particle position. In addition, no clear difference is observed between the voids marked by the blue and green frames due to the same particle position within the void.

In addition, we extended the polarization-dependent image analysis from a single linear polarizer to two linear polarizers in a crossed configuration. Figure 6a displays an optical microscope image acquired using crossed polarizers, with the corresponding SEM image shown in Figure 6b. Comparison of the optical and SEM images reveals that the particles occupy different positions within the individual voids. The voids have a diameter of approximately 560 nm and a depth of about 200 nm. In this experiment, the void dimensions are deliberately mismatched with respect to the deposited particles, which have a nominal diameter of 300 nm, to allow different particle positions within the voids. Although such states would typically be removed during the cleaning procedure, careful cleaning using low ultrasonication power made it possible to preserve them. Comparing the magnified views depicted in Figure 6c, the different particle positions within the voids can already be inferred from the spatial color distribution in the optical microscope image. The colored frames indicate the corresponding voids in the full optical microscope image. Consistently, the blue- and green-framed images show particles located at nearly the same position within the voids, resulting in a similar spatial color distribution in the optical microscope images.

### Simultaneous morphology- and material-sensitive nanoplastic detection

To consolidate the shape- and material-sensitive capabilities of the detection strip, we perform a final experiment in which both parameters are probed simultaneously. This measurement is designed to demonstrate concurrent discrimination of particle shape and material within a single experiment.
Ellipsoidal PS particles with an AR of 2:1 are co-deposited with spherical particles composed of PS, PMMA, and PET, each with a nominal diameter of 300 nm. The relatively small AR of the ellipsoidal particles was deliberately chosen to increase the experimental challenge, as these ellipses approach spherical symmetry more closely than particles with larger ARs. This experimental configuration results in six distinct cases, as schematically illustrated in Figure 7a. DLS measurements performed for the PET particles yield an average diameter of 303 nm with a PDI of 0.084 . Among the investigated materials, PET exhibits the highest refractive index ($n = 1.64$ @ 633 nm)[54].
A SEM image of the deposited sample prior to cleaning can be seen in Figure 7b. Both ellipsoidal and spherical particles of the three different materials are clearly visible. However, the individual material types of the spherical particles cannot be distinguished based on morphology alone.
Figure 7c displays a representative section of the detection strip comprising $5 \times 20$ fields. Each field contains four elliptical voids (left) and four circular voids (right). The circular voids have a diameter of 330 nm, whereas the elliptical voids possess major and minor axis lengths of 610 nm and 260 nm, respectively. The depth of the voids is approximately 280 nm, and the center-to-center distance is 1.2 μm. Magnified views of selected regions (Figure 7d,e) reveal a pronounced color variation that depends on whether the voids are occupied by ellipsoidal particles or by spherical particles of different materials.
The average color of each void is determined using automated color extraction and plotted in the CIE chromaticity diagram. The material sensitivity of the platform is summarized in Figure 7f, where the colors associated with the circular voids form four clearly separated clusters corresponding to empty voids and to voids filled with PS, PMMA, or PET particles. This assignment is supported by simulated reflection spectra and corresponding colors for the three materials and the empty case (Supplementary Fig. 16). Please note that deviations between the simulated and experimentally observed colors can arise due to experimental factors such as variations in void geometries, particle size distributions, effective refractive indices, and differences in the applied white balance. Moreover, reference measurements were performed with only a single particle type for PS (Supplementary Fig. 17), PMMA (Supplementary Fig. 18), and PET (Supplementary Fig. 19). A second set of reference measurements was employed to ensure the highest reliability of the material discrimination.
Figure 7g presents the CIE diagram for the elliptical voids, where a clear distinction between filled and empty states is observed. The colors of elliptical and circular voids are therefore analyzed separately, as empty elliptical voids exhibit colors similar to circular voids filled with PMMA particles.

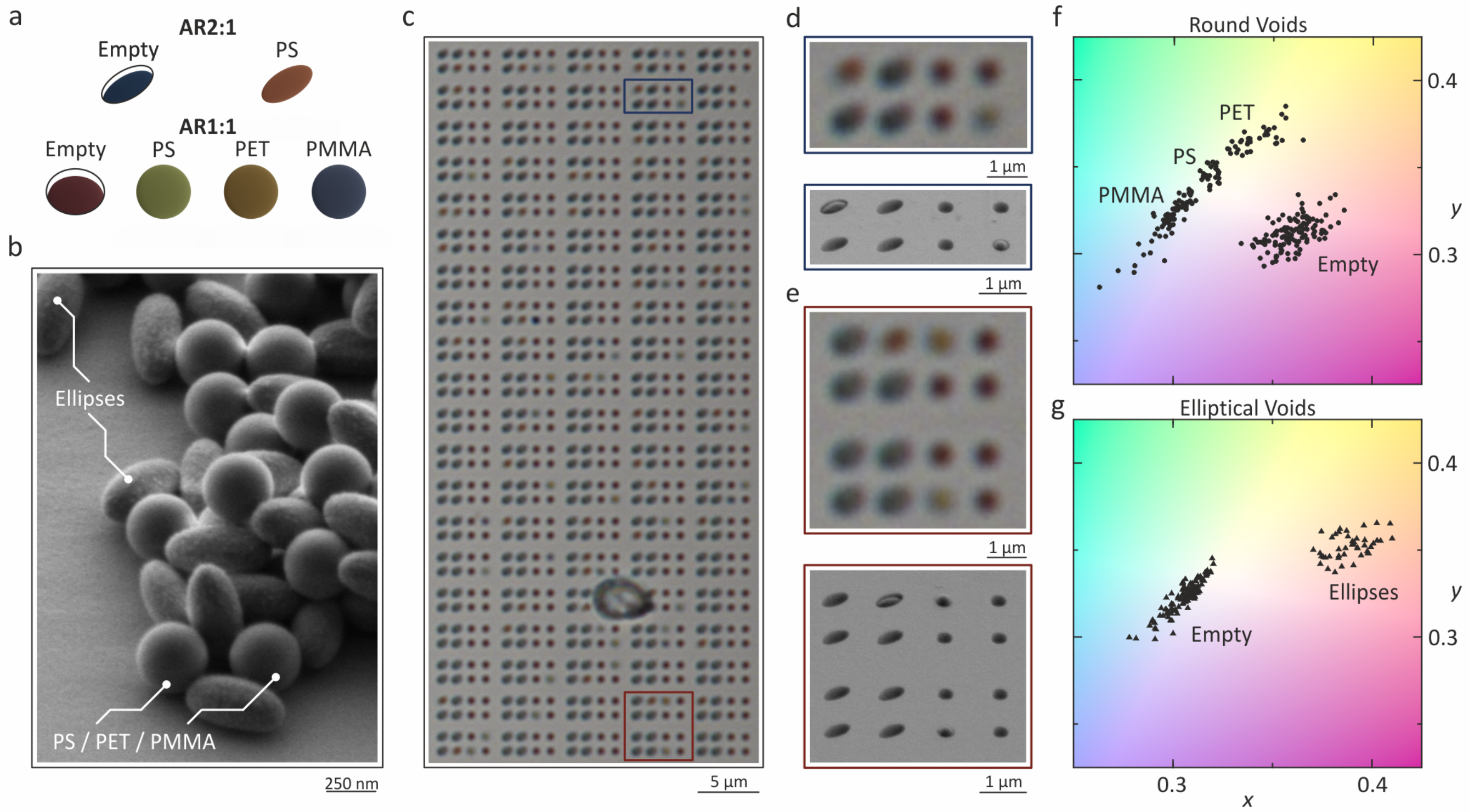

**Figure 7: Simultaneous morphology- and material-sensitive nanoplastic detection. a**, Deposited mixture consisting of ellipsoidal particles with an aspect ratio of 2:1 and round particles of three different polymer materials. **b**, SEM image of this mixed nanoparticle sample. **c**, Optical microscope image of the full detection strip. **d & e**, Magnified microscope images (top) of selected void areas with corresponding SEM images (bottom). Ellipsoidal particles are selectively trapped in elliptical voids, while spherical particles only occupy round voids. **f & g**, Color representation in CIE-1931 diagrams for spherical voids and elliptical voids. For the spherical voids, four color clusters can be observed, indicating the three different materials as well as the empty voids. For the elliptical voids, the empty and filled voids can be distinguished.

## Conclusion

In summary, we introduce a sensing platform that enables the simultaneous, shape- and material-sensitive detection of individual nanoplastic particles with diameters down to 300 nm, based on the optical response of Mie void resonators. By exploiting Mie-type resonances in nanoscale void structures embedded within a high-refractive-index material, nanoparticles are selectively trapped and reported through a refractive-index-dependent color signature that is directly accessible using a standard optical microscope. Carefully designed void geometries allow reliable shape sorting of spherical and ellipsoidal particles with different aspect ratios, while polymer nanoparticles of comparable size but distinct refractive indices, including PS, PMMA, and PET, are unambiguously discriminated via their characteristic resonance colors. This capability extends to complex particle mixtures, enabling the concurrent extraction of morphological and compositional information within a *single* measurement.
Detection fidelity is further enhanced through polarization-resolved imaging and differential color analysis, which increase sensitivity to subtle refractive-index perturbations. Beyond its analytical performance, a key strength of the presented approach lies in its exceptional simplicity: the readout of the platform relies exclusively on widely available optical microscopy and automated image analysis, without the need for complex spectroscopy or elaborate instrumentation. In combination with complementary techniques such as μ-FTIR, μ- and stimulated Raman spectroscopy, the trapped particles remain accessible for downstream, high-specificity chemical analysis.
Mie-void-based optical detection thus represents a versatile and powerful addition to the nanoplastic characterization toolbox, bridging the gap between laboratory-grade analysis and practical, environmental monitoring. Future research may focus on the simultaneous optimization of void geometries to both match the deposited particles and enhance the optical resonances. In addition, surface functionalization could enable operation in more complex environments and further improve the device. Finally, the system performance could be validated using real environmental and biological samples, such as blood specimens. Ultimately, we envision this platform enabling simple, rapid, and field-deployable nanoplastic test strips and mobile detection devices, opening new avenues for large-scale screening of nanoscale plastic pollution across diverse settings.

## Acknowledgments

The authors thank Jonas Herbig for his support with photographs and image analysis, and Bekim Hisenaj for valuable input on data analysis based on principal component analysis. The authors also acknowledge Prof. Anke Krueger and Dr. Johannes Ackermann for their support with dynamic light scattering measurements.
We acknowledge a grant through the Baden-Wuerttemberg-Stiftung (Opterial), Ministerium für Wissenschaft, Forschung und Kunst Baden-Württemberg (RiSC, Innovation Campus Future Mobility; Sd Manu1, Lab7), Vector Stiftung (MINT Innovations), European Research Council (Advanced Grant Complex-plas, PoC Grant 3DPrintedOptics), the German Research Foundation (SPP1839 Tailored Disorder; grant number 431314977/GRK2642), Bundesministerium für Bildung und Forschung (Printoptics, Q.Link.X, QR.X), HORIZON EUROPE European Innovation Council (IV-LAB; grant number 101115-545), Carl-Zeiss-Stiftung (CZS Center QPhoton, Endo-Print3D) and the University of Stuttgart (Terra Incognita).
M.R. acknowledges funding and support from the International Research Training Group (IRTG) OPTEXC (project number 464648186) (German Research Foundation).

## Author Contribution Statement

D.L., S.A., M.U., and M.H. performed the measurements and contributed to the definition of the experimental procedure. M.H. conceived the experiments. M.R., H.G. and M.H. supervised the project. D.L. and M.H. evaluated and analysed the results. J.S. designed and performed all simulations. E.K. and M.R. expanded the paper by providing additional material. All authors participated in the generation of the paper and contributed to the final version of the work.

## Competing Interests Statement

The authors declare no competing interests.

## Correspondence

Correspondence and requests for materials should be addressed to M. Hentschel.

## References

1. MacLeod, M., Arp, H. P. H., Tekman, M. B. & Jahnke, A. The global threat from plastic pollution. *Science* **373,** 61–65 (2021).
2. Thompson, R. C. *et al.* Twenty years of microplastic pollution research - what have we learned? *Science* **386,** 395 (2024).
3. Fusco, L. *et al.* Nanoplastics: Immune Impact, Detection, and Internalization after Human Blood Exposure by Single-Cell Mass Cytometry. *Adv. Mater.* **37,** 2413413 (2025).
4. Silva, A. B. *et al.* Microplastics in the environment: Challenges in analytical chemistry - A review. *Anal. Chim. Acta* **1017,** 1–19 (2018).
5. Lebreton, L. *et al.* Evidence that the Great Pacific Garbage Patch is rapidly accumulating plastic. *Sci. Rep.* **8,** 4666 (2018).
6. Carpenter, E. J. & Smith Jr., K. L. Plastics on the Sargasso Sea Surface. *Science* **175,** 1240–1241 (1972).
7. Carpenter, E. J., Anderson, S. J., Harvey, G. R., Miklas, H. P. & Peck, B. B. Polystyrene Spherules in Coastal Waters. *Science* **178,** 749–750 (1972).
8. Abdeljaoued, A. *et al.* Efficient removal of nanoplastics from industrial wastewater through synergetic electrophoretic deposition and particle stabilized foam formation. *Nat. Commun.* **15,** 5437 (2024).
9. Ambrosini, R. *et al.* First evidence of microplastic contamination in the supraglacial debris of an alpine glacier. *Environ. Pollut.* **253,** 297–301 (2019).
10. Scheurer, M. & Bigalke, M. Microplastics in Swiss Floodplain Soils. *Environ. Sci. Technol.* **52,** 3591–3598 (2018).
11. Ostle, C. *et al.* The rise in ocean plastics evidenced from a 60-year time series. *Nat. Commun.* **10,** 1622 (2019).
12. Peng, X. *et al.* Microplastics contaminate the deepest part of the world's ocean. *Geochem. Perspect. Lett.* **9,** 1–5 (2018).
13. Russell, M. & Webster, L. Microplastics in sea surface waters around Scotland. *Mar. Pollut. Bull.* **166,** 112210 (2021).
14. Prata, J. C., da Costa, J. P., Duarte, A. C. & Rocha-Santos, T. Methods for sampling and detection of microplastics in water and sediment: A critical review. *TrAC* **110,** 150–159 (2019).
15. Lau, W. W. Y. *et al.* Evaluating scenarios toward zero plastic pollution. *Science* **369,** 1455–1461 (2020).
16. Cai, H. *et al.* Analysis of environmental nanoplastics: Progress and challenges. *Chem. Eng. J.* **410,** 128208 (2021).
17. da Costa, J. P., Santos, P. S., Duarte, A. C. & Rocha-Santos, T. (Nano)plastics in the environment - Sources, fates and effects. *Sci. Total Environ.* **566-567,** 15–26 (2016).
18. Hartmann, N. B. *et al.* Are We Speaking the Same Language? Recommendations for a Definition and Categorization Framework for Plastic Debris. *Environ. Sci. Technol.* **53,** 1039–1047 (2019).

19. Eerkes-Medrano, D., Leslie, H. A. & Quinn, B. Microplastics in drinking water: A review and assessment. *Curr. Opin. Environ. Sci. Health* **7,** 69–75 (2019).
20. Singh, S., Trushna, T., Kalyanasundaram, M., Tamhankar, A. J. & Diwan, V. Microplastics in drinking water: a macro issue. *Water Supply* **22,** 5650–5674 (2022).
21. Cverenkárová, K., Valachovičová, M., Mackuľak, T., Žemlička, L. & Bírošová, L. Microplastics in the food chain. *Life* **11,** 1349 (2021).
22. Cole, M. & Galloway, T. S. Ingestion of Nanoplastics and Microplastics by Pacific Oyster Larvae. *Environ. Sci. Technol.* **49,** 14625–14632 (2015).
23. Galloway, T. S., Cole, M. & Lewis, C. Interactions of microplastic debris throughout the marine ecosystem. *Nat. Ecol. Evol.* **1,** 0116 (2017).
24. Truong, T. N. S. *et al.* Microplastic in atmospheric fallouts of a developing Southeast Asian megacity under tropical climate. *Chemosphere* **272,** 129874 (2021).
25. Zhang, Y. *et al.* Atmospheric microplastics: A review on current status and perspectives. *Earth-Sci. Rev.* **203,** 103118 (2020).
26. Prata, J. C., da Costa, J. P., Lopes, I., Duarte, A. C. & Rocha-Santos, T. Environmental exposure to microplastics: An overview on possible human health effects. *Sci. Total Environ.* **702,** 134455 (2020).
27. Zhu, L. *et al.* Transport of microplastics in the body and interaction with biological barriers, and controlling of microplastics pollution. *Ecotoxicol. Environ. Saf.* **255,** 114818 (2023).
28. Zhao, X. *et al.* Defining the size ranges of polystyrene nanoplastics according to their ability to cross biological barriers. *Environ. Sci. Nano* **10,** 2634–2645 (2023).
29. Pozo, K. *et al.* Persistent organic pollutants sorbed in plastic resin pellet - "Nurdles" from coastal areas of Central Chile. *Mar. Pollut. Bull.* **151,** 110786 (2020).
30. Urso, M., Ussia, M., Nowotný, F. & Pumera, M. Trapping and detecting nanoplastics by MXene-derived oxide microrobots. *Nat. Commun.* **13,** 3573 (2022).
31. Zhang, Y. *et al.* State-of-the-art evolution of detection, adsorption, and degradation of micro/nano-plastics. *J. Mater. Sci. Technol.* **263,** 97–116 (2026).
32. Daoutakou, M. & Kintzios, S. Biosensors for Micro- and Nanoplastic Detection: A Review. *Chemosensors* **13,** 143 (2025).
33. Krauss, T. F., Miller, L., Wälti, C. & Johnson, S. Photonic and electrochemical biosensors for near-patient tests - a critical comparison. *Optica* **11,** 1408–1418 (2024).
34. Yesilkoy, F. *et al.* Ultrasensitive hyperspectral imaging and biodetection enabled by dielectric metasurfaces. *Nat. Photon.* **13,** 390–396 (2019).
35. Santos, F. A., Andre, R. S., Alvarenga, A. D., Alves, A. L. M. M. & Correa, D. S. Micro- and nanoplastics in the environment: a comprehensive review on detection techniques. *Environ. Sci.: Nano* **12,** 3442–3467 (2025).
36. Kalaronis, D. *et al.* Microscopic techniques as means for the determination of microplastics and nanoplastics in the aquatic environment: A concise review. *Green Analytical Chemistry* **3,** 100036 (2022).
37. Altug, H., Oh, S.-H., Maier, S. A. & Homola, J. Advances and applications of nanophotonic biosensors. *Nat. Nanotechnol.* **17,** 5–16 (2022).

38. El-Helou, A. J. *et al.* Optical Metasurfaces for the Next-Generation Biosensing and Bioimaging. *Laser Photonics Rev.* **19,** 2401715 (2025).

39. Cabernard, L., Roscher, L., Lorenz, C., Gerdts, G. & Primpke, S. Comparison of Raman and Fourier Transform Infrared Spectroscopy for the Quantification of Microplastics in the Aquatic Environment. *Environ. Sci. Technol.* **52,** 13279–13288 (2018).

40. Ardini, B. *et al.* Fast Detection and Classification of Microplastics by a Wide-Field Fourier Transform Raman Microscope. *Environ. Sci. Technol.* **59,** 9255–9264 (2025).

41. Yi, J. *et al.* Surface-enhanced Raman spectroscopy: a half-century historical perspective. *Chem. Soc. Rev.* **54,** 1453–1551 (2025).

42. Sánchez-Alvarado, A. B. *et al.* Combined Surface-Enhanced Raman and Infrared Absorption Spectroscopies for Streamlined Chemical Detection of Polycyclic Aromatic Hydrocarbon-Derived Compounds. *ACS Nano* **17,** 25697–25706 (2023).

43. Xu, G. *et al.* Surface-Enhanced Raman Spectroscopy Facilitates the Detection of Microplastics <1 µm in the Environment. *Environ. Sci. Technol.* **54,** 15594–15603 (2020).

44. Kniggendorf, A.-K., Wetzel, C. & Roth, B. Microplastics Detection in Streaming Tap Water with Raman Spectroscopy. *Sensors* **19,** 1839 (2019).

45. Araujo, C. F., Nolasco, M. M., Ribeiro, A. M. & Ribeiro-Claro, P. J. Identification of microplastics using Raman spectroscopy: Latest developments and future prospects. *Water Res.* **142,** 426–440 (2018).

46. Hentschel, M. *et al.* Dielectric Mie voids: confining light in air. *Light Sci. Appl.* **12,** 3 (2023).

47. Sarbajna, A. *et al.* Encapsulated Void Resonators in Van der Waals Heterostructures. *Laser Photonics Rev.* **19,** 2401215 (2025).

48. Hamidi, M. *et al.* Quasi-Babinet principle in dielectric resonators and Mie voids. *Phys. Rev. Res.* **7,** 013136 (2025).

49. Ludescher, D. *et al.* Optical sieve for nanoplastic detection, sizing and counting. *Nat. Photon.* **19,** 1138–1145 (2025).

50. Krauss, T. F. Mie resonances light up nanoplastics. *Nat. Photon.* **19,** 1031, 1032 (2025).

51. Lal, S., Link, S. & Halas, N. J. Nano-optics from sensing to waveguiding. *Nat. Photon.* **1,** 641–648 (2007).

52. Arslan, S. *et al.* Attoliter Mie Void Sensing. *ACS Photonics* **12,** 3950–3958 (2025).

53. Sultanova, N., Kasarova, S. & Nikolov, I. Dispersion Properties of Optical Polymers. *APPA* **116,** 585–587 (2009).

54. Elman, J., Greener, J., Herzinger, C. & Johs, B. Characterization of biaxially-stretched plastic films by generalized ellipsometry. *Thin Solid Films* **313-314,** 814–818 (1998).

55. Benke, D. *et al.* Prolate spheroidal polystyrene nanoparticles: matrix assisted synthesis, interface properties, and scattering analysis. *Soft Matter* **19,** 9006–9016 (2023).

56. Champion, J. A., Katare, Y. K. & Mitragotri, S. Making polymeric micro- and nanoparticles of complex shapes. *Proc. Natl. Acad. Sci.* **104,** 11901–11904 (2007).

57. Aspnes, D. E., Kelso, S. M., Logan, R. A. & Bhat, R. Optical properties of AlxGa1-x As. *J. Appl. Phys.* **60,** 754–767 (1986).

58. Schwab, J. *et al.* Illuminating Mie Voids: An Analytical Model for Nanophotonics. *ACS Photonics* (in press, 2026).

## Methods

*Mie Void Sample Patterning*

The Mie voids used for the nanoplastic detection are fabricated in gallium arsenide (GaAs) wafers using a dry-etching approach optimized for rapid and reliable pattern transfer. The combination of electron-beam lithography (EBL) and plasma etching enables a straightforward fabrication sequence compatible with cost-effective, single-use test structures.

The patterning on pre-cleaned GaAs substrates starts with spin-coating a 400 nm electron-beam resist (AR-P 6200.13). The resist is coated at 2000 rpm for 5 s, followed by 4000 rpm for 55 s. Subsequently, it is soft-baked at 180 °C for 3 min. Patterns are defined using an EBL system (VOYAGER, Raith GmbH) operated at 50 kV acceleration voltage with a 60 µm aperture, 10 nm step size, and an exposure dose of 200 $\mu C/cm^2$.

Following the exposure, the samples are developed in AR 600-546 for 90 s, immersed in AR 600-60 stopper solution for 30 s, and rinsed in isopropanol for 30 s. The pattern transfer into GaAs is then carried out using an inductively coupled plasma system (Plasmalab System 100, Oxford Instruments) with $SiCl_4$ gas. Etching is performed at 6 W radio-frequency power and 350 W inductively coupled plasma power, with gas flows of 6 sccm $SiCl_4$ and 5 sccm helium at a chamber temperature of 30 °C. Etching durations range from 30 s to 1 min, depending on the target void depth.

Post-etching, the residual resist mask is removed by brief immersion in an N-ethylpentedrone solution to ensure complete lift-off. Samples are then rinsed in acetone and isopropanol and dried under nitrogen.

*Nanoplastic Particles*

In this study, both spherical and ellipsoidal nanoplastic particles are investigated. Spherical particles made of polystyrene (PS), poly(methyl methacrylate) (PMMA), and polyethylene terephthalate (PET) with a nominal diameter of 300 nm are purchased from Abvigen Inc, US. Note that the PET particles are prepared based on PS spheres as seeds. Consequently, their composition only contains at least 75 % PET, yielding a refractive index slightly lower than that of pure PET.

Ellipsoidal particles were fabricated using a scalable deformation process that has been reported by Benke et al. and others[55,56]. Monodisperse PS spheres (350 nm diameter) are embedded in a polyvinylalcohol matrix (Mowiol. 8-88, Mw = 67000 , 86.7 - 88.7 % hydrolysis, Merck/Sigma-Aldrich) cast as thin films and subsequently mounted in a mechanical stretching device. After heating the films above the glass-transition temperature of PS (150 °C), the particles are anisotropically deformed through incremental stretching steps[56]. The final particle aspect ratio is controlled by the total applied strain: a strain of 141 % yielded an aspect ratio of 2:1, while strains of 171 % and 210 % produced aspect ratios of 3:1 and 4:1, respectively. Following deformation, the polyvinylalcohol matrix is dissolved, and the ellipsoidal particles are isolated by centrifugation. All particle batches are stored as electrostatically stabilized aqueous dispersions.

Prior to particle deposition, all detection samples are treated in oxygen plasma to increase surface hydrophilicity. After deposition and complete drying, loosely bound particles are removed by immersing the samples in water and applying ultrasonication. Ultrasonic cleaning is performed for 2 min at 35 kHz, with applied voltages between 50 V and 260 V to power the ultrasonication depending on the complexity of the deposited sample matrix. During this cleaning procedure, mismatched particles are removed when the interaction forces are insufficient to retain them within the voids. Consequently, this cleaning acts as a filtering step that translates the degree of matching between particle and void dimensions into the observed trapping efficiency. This relationship enables the precise determination of particle size and shape using Mie voids.

*Microscopic Imaging*
Optical images of empty and filled voids are acquired in reflection and bright-field modes using a Nikon Eclipse LV100NM microscope. Images are captured using objectives with magnifications of 20× (NA 0.45), 50× (NA 0.8), and 100× (NA 0.9). To reduce the effective numerical aperture, the aperture stop in the illumination path is fully closed. Color accuracy is ensured by performing white-balance calibration on a clean, blank GaAs substrate prior to each image acquisition, and by automatically optimizing the camera exposure time. Illumination is provided by a Nikon halogen lamp (LV-LH50PC), and images are recorded using NIS Elements D software paired with a CCD camera. For polarization-dependent imaging with a single polarizer (Figures 3, 4, and 5), a rotatable linear polarizer is inserted into the illumination path. For crossed-polarizer imaging, a second, fixed linear polarizer is additionally inserted into the detection path (Figure 6).
Scanning electron microscopy (SEM) images of the samples are obtained using either a Hitachi S-4800 or a Zeiss Gemini 560 SEM. Images are typically acquired at a 30° or 45° observation angle, and the acceleration voltage is adjusted depending on the sample type. When examining samples containing PMMA or PET nanoparticles, reduced acceleration voltages are used to prevent beam-induced particle shrinkage or deformation during imaging.

*Color Readout*
A MATLAB script is used to extract and analyze color information from the optical microscopy images. The script automatically identifies all individual voids, whether empty or particle-filled, based on their periodic arrangement on the surface and their color contrast relative to the GaAs substrate. After detection, the user specifies the void radius in pixels, and the script computes the mean RGB values within each void area. Each void is thus represented by a three-channel RGB vector corresponding to its red, green, and blue intensities. The extracted color values are subsequently mapped either onto a CIE 1931 chromaticity diagram (Figure 2, Figure 5, and Figure 7) or into an RGB color cube (Figure 3). These representations highlight subtle color variations and enable reliable differentiation between empty voids and voids filled with nanoplastic particles of different materials.

*Dynamic Light Scattering*
Particle size distributions of PS, PMMA, and PET nanoparticles were measured using dynamic light scattering (Zetasizer Nano ZS, Malvern, USA). The particles all had a nominal diameter of 300 nm. Prior to measurements, the suspensions were diluted in distilled water, and approximate refractive indices for each polymer were used as input parameters[53,54]. For each material, three measurements were acquired and averaged.
In Supplementary Fig. 2, the particle diameter is presented as a volume distribution. The mean particle diameters obtained from DLS are 363 nm (PS), 374 nm (PMMA), and 303 nm (PET). Moreover the polydispersity index (PDI) for the specific materials is given by 0.099 (PS), 0.005 (PMMA), and 0.084 (PET).

*Simulations*
Reflectance spectra and the corresponding color appearances (Supplementary Fig. 16) for empty and nanoplastic-filled voids (PS, PMMA, and PET) are computed using the frequency-domain finite-element solver in COMSOL Multiphysics. The structures are modeled as conical voids etched into a GaAs substrate, with air ($n = 1$) occupying the void volume and the superstrate region. Hole dimensions are extracted from SEM images of the fabricated samples. For simulations including nanoplastics, spherical particles of the corresponding material are placed inside the voids. Refractive indices for GaAs, PS, PMMA, and PET are taken from the literature[53,54,57]. Please note that, due to fabrication constraints, the PET particles consist of at least 75 % PET, while the remaining fraction is formed by a PS core. This composite structure influences the effective refractive index of the particles. However, since the exact composition is not known, this effect is not explicitly considered in the simulations.
To approximate an infinitely extended system along the optical axis, perfectly matched layers (PMLs) are placed at the top and bottom boundaries of the domain. Laterally, the system is treated as periodic, with a period of 460 nm for spheres with diameters of 300 nm. Periodic boundary conditions are applied accordingly. Illumination is introduced through a periodic port in the air region above the structure, launching an x-polarized plane wave propagating in the negative z-direction. In contrast to the experimental procedure, where the white balance is calibrated using the reflection of the bare GaAs substrate, the simulations use an ideal white reference for the balancing. An improved analytical model for predicting the optical color response was recently published [58].

# - Supplementary Information -
Size, Shape, and Material matter: All-optical Mie void sensor for complex nanoplastic mixtures

D. Ludescher[1,*], J. Schwab[1], S. Arslan[1], E. Kubacki[2], M. Ubl[1], M. Retsch[2,3], H. Giessen[1], and M. Hentschel[1,*]

[1] *4th Physics Institute and Research Center SCoPE, University of Stuttgart, 70569 Stuttgart, Germany*
[2] *Department of Chemistry, Physical Chemistry I, University of Bayreuth, 95447 Bayreuth, Germany*
[3] *Bavarian Polymer Institute, Bayreuth Center for Colloids and Interfaces, Bayreuther Institut für Makromolekülforschung, and Bavarian Center for Battery Technology (BayBatt), University of Bayreuth, 95447 Bayreuth, Germany*
[*] *Corresponding authors: Mario Hentschel (m.hentschel@pi4.uni-stuttgart.de) and Dominik Ludescher (d.ludescher@pi4.uni-stuttgart.de)*

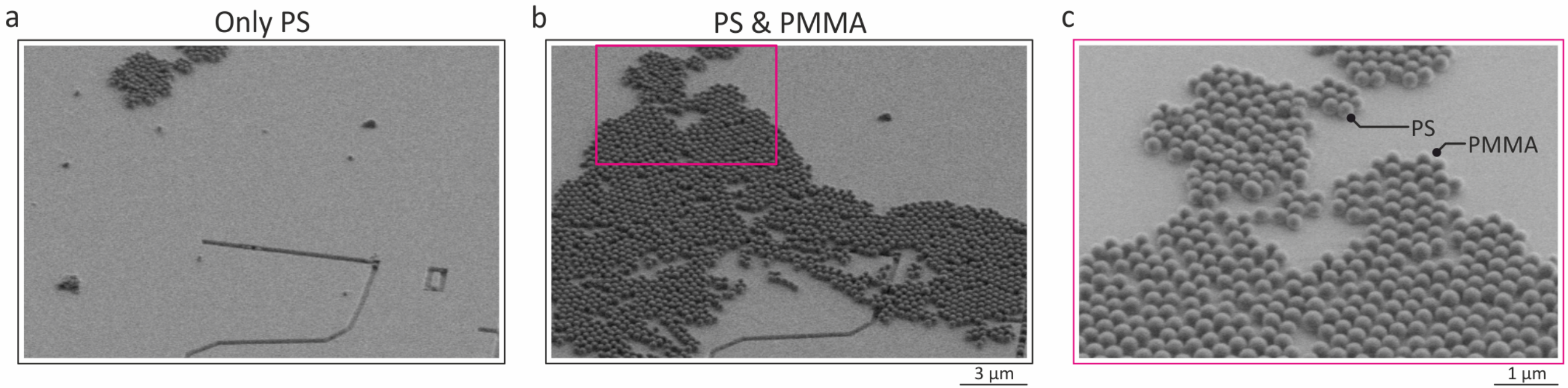

**SI 1: Comparison of PS and PMMA particles.** SEM images are used to compare PS and PMMA particles. **a**, SEM image only showing PS particles, forming a cluster in the top-left region. **b**, SEM image after additional deposition of PMMA particles on the same sample, without removing the initial PS particles. **c**, Magnified view of the marked pink area, where PS and PMMA particles are located adjacent to each other. No distinct optical contrast between the two particle types is observed in the SEM images.

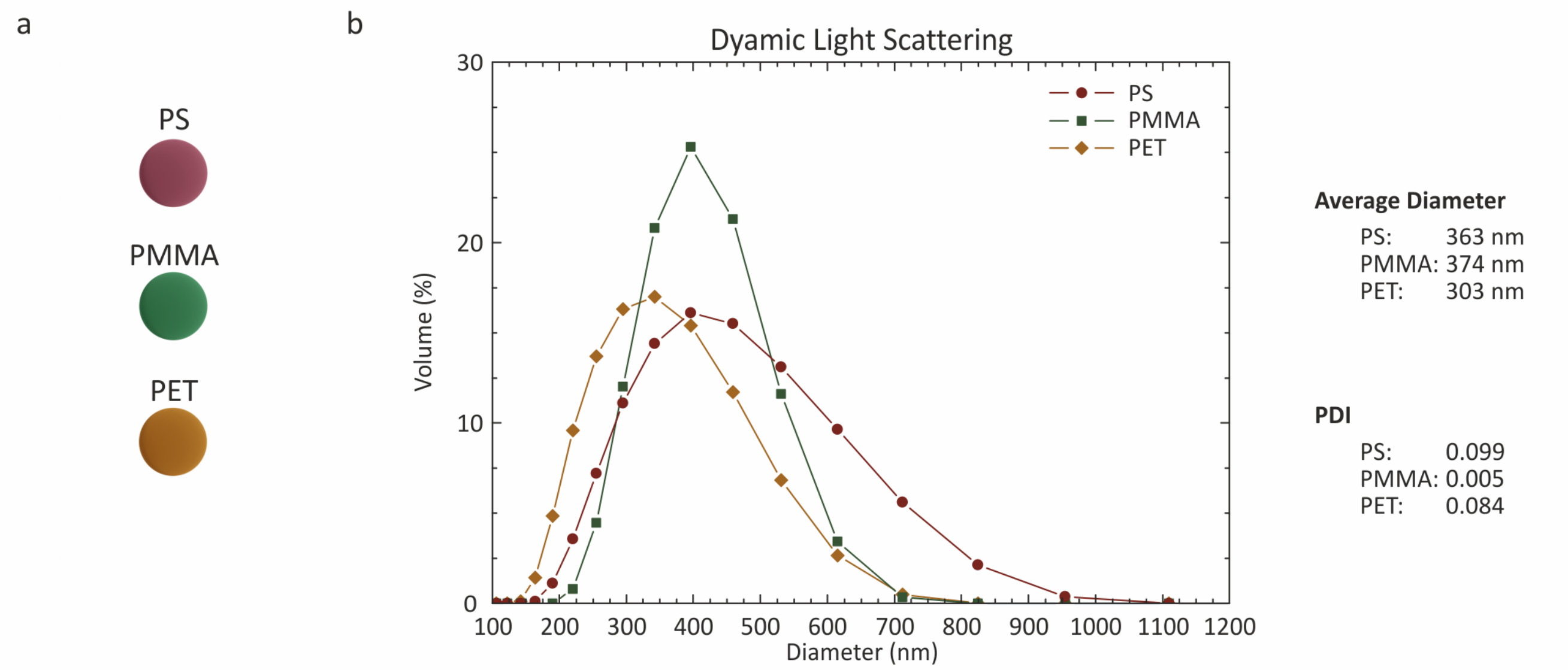

**SI 2: Dynamic light scattering measurements for different particle materials.** **a**, Measured particle size distributions obtained by DLS for PS, PMMA, and PET nanoparticles. **b**, Volume percentage as a function of particle diameter, with the corresponding average diameter and polydispersity index (PDI) indicated for each material.

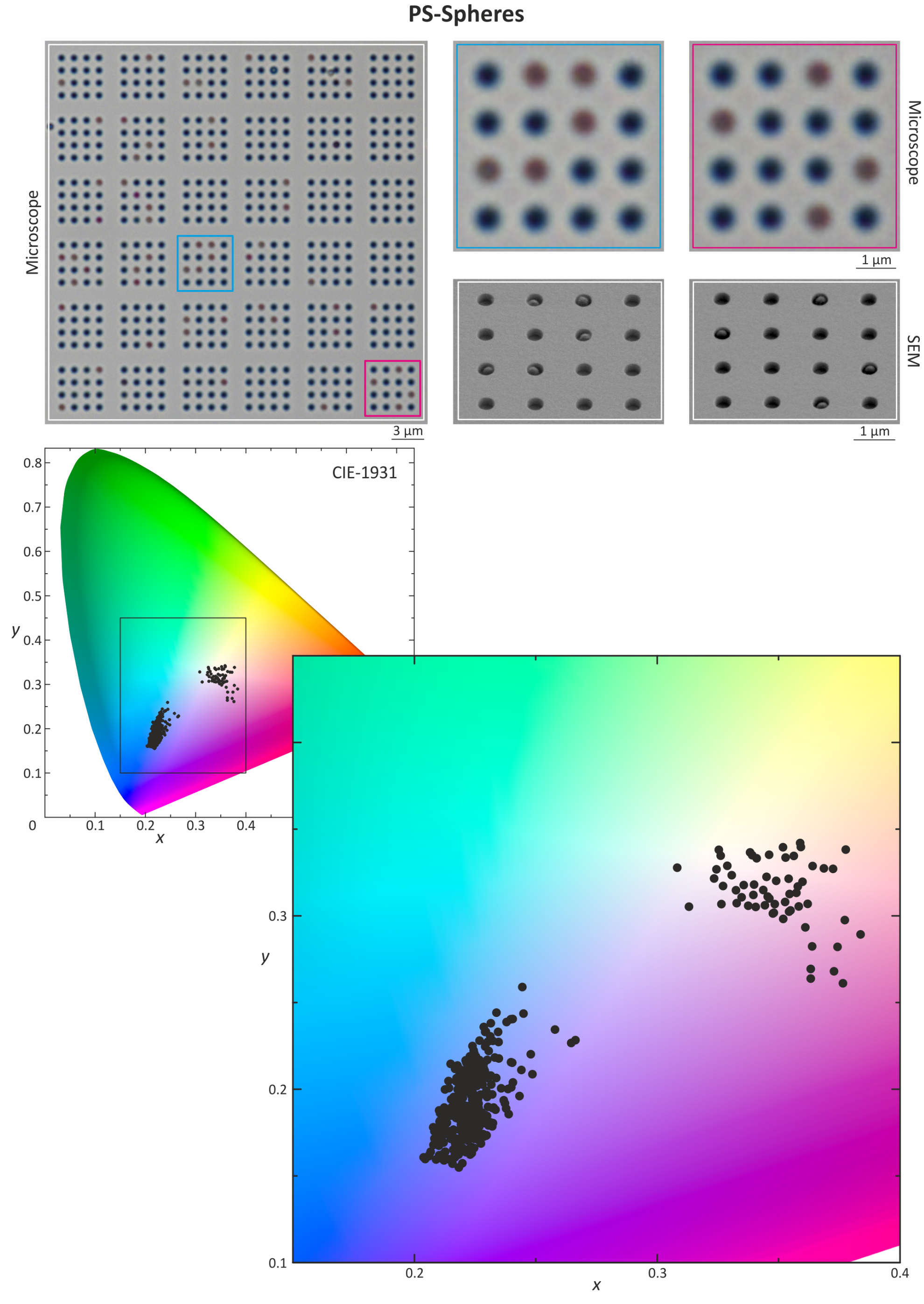

**SI 3: Color reference measurement for PS particles.** PS particles are deposited to determine the characteristic color of voids filled with PS. (Top) Optical image of the detection strip with two magnified regions and corresponding SEM images. (Bottom) Full CIE-1931 chromaticity diagram and a magnified view of the relevant region, indicating a red color associated with PS-filled voids.

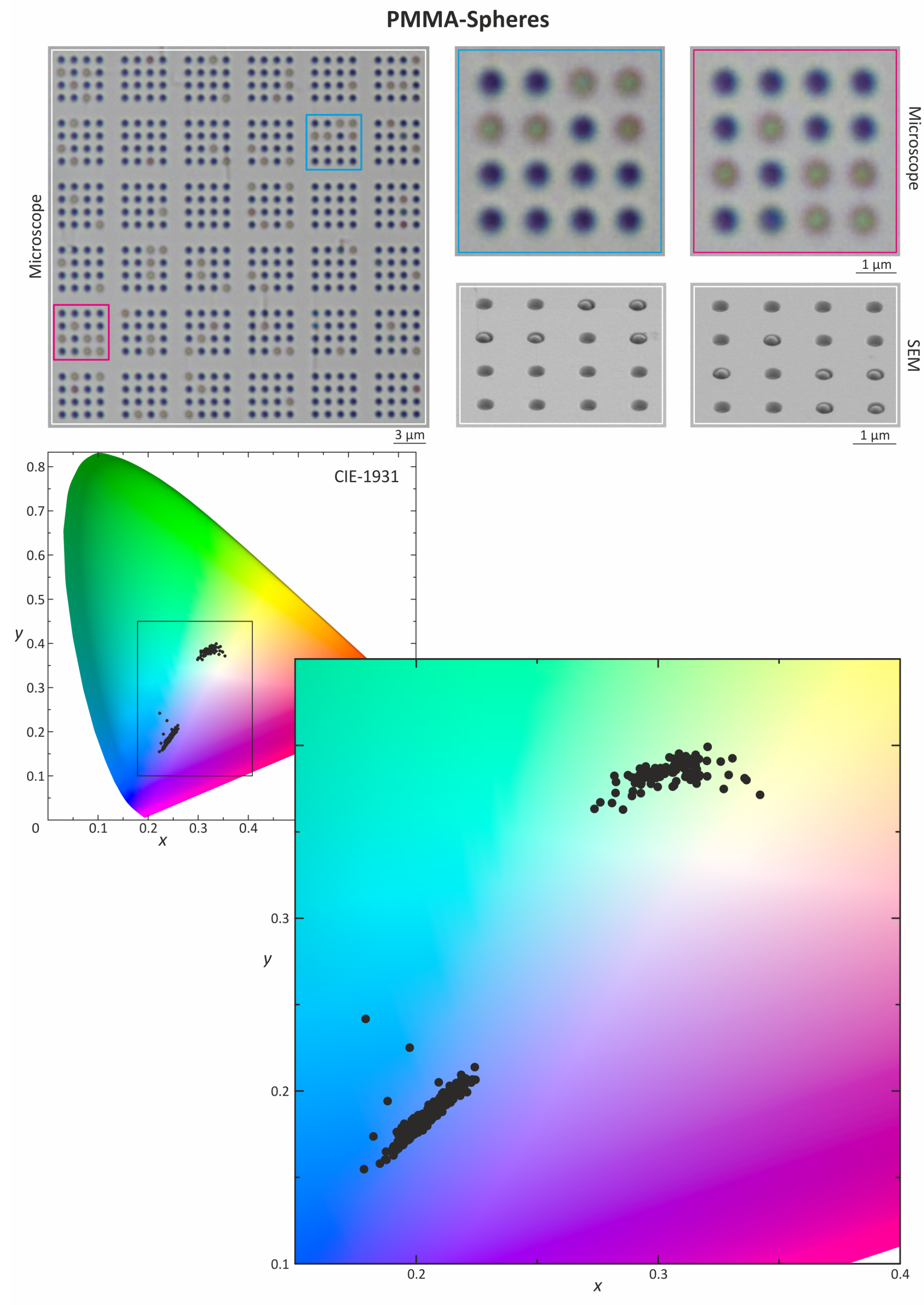

**SI 4: Color reference measurement for PMMA particles.** PMMA particles are deposited to determine the characteristic color of voids filled with PMMA. (Top) Optical image of the detection strip with two magnified regions and corresponding SEM images. (Bottom) Full CIE-1931 chromaticity diagram and a magnified view of the relevant region, indicating a yellow or green color associated with PMMA-filled voids.

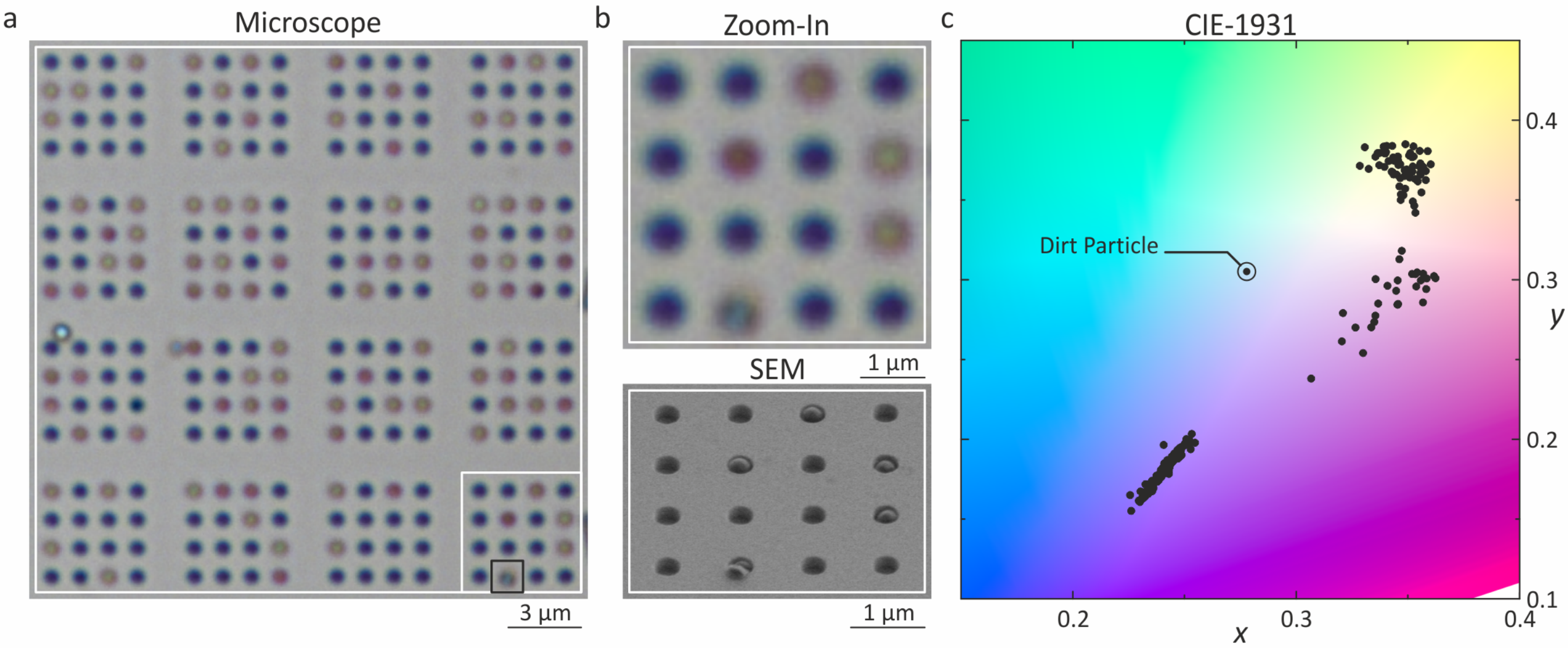

**SI 5: Investigation of the influence of dirt particles. a**, Microscope image of a detection strip containing a dirt particle in the lower-right corner. The void filled with the dirt particle is indicated by the black square. **b**, Magnified view of the same region with the corresponding SEM image showing the dirt particle. **c**, CIE-1931 color representation of all measured colors, revealing one clear outlier corresponding to the void filled with the dirt particle.

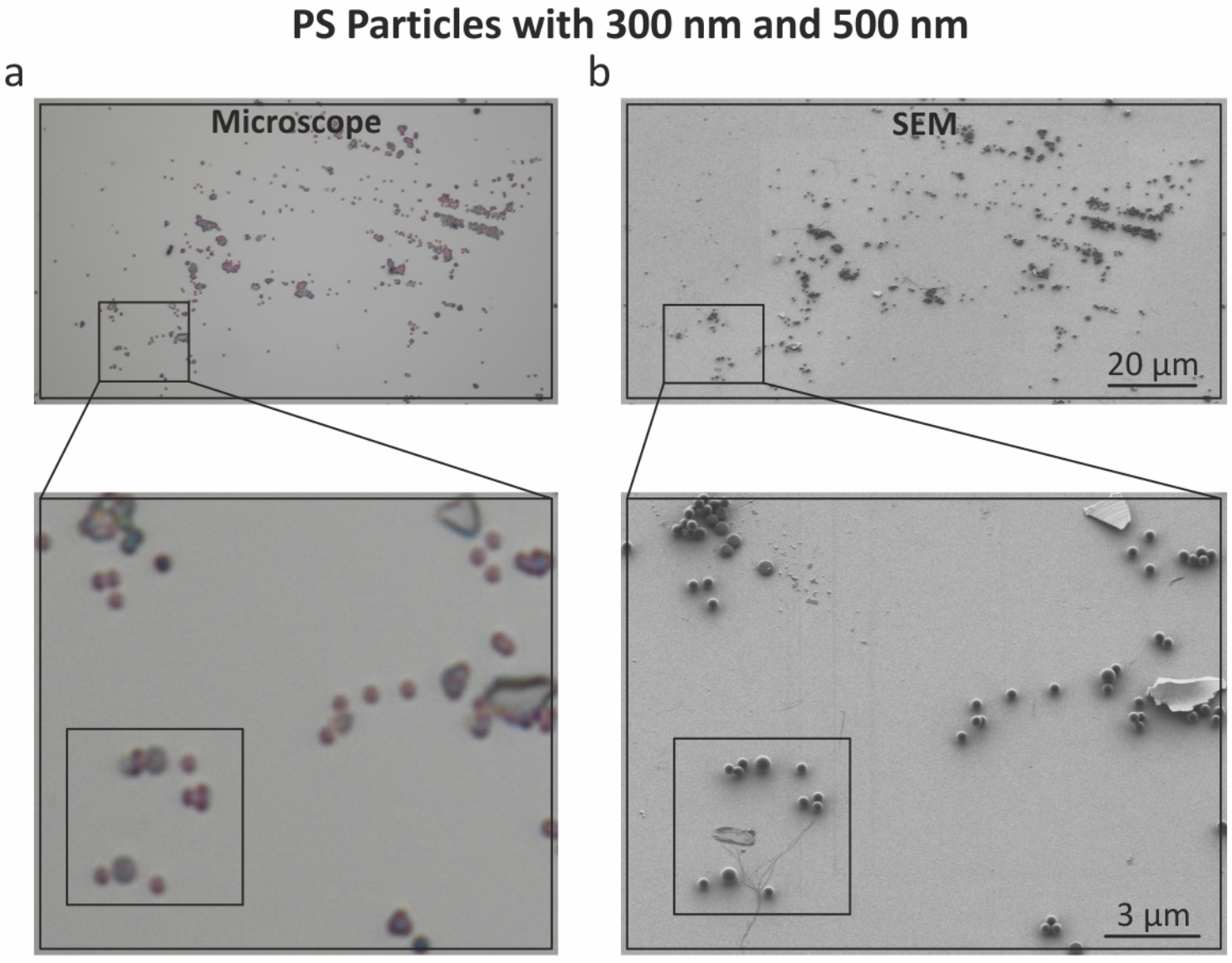

**SI 6: PS particles with different diameters deposited on a flat GaAs substrate. a**, Optical microscope image of PS particles with nominal diameters of 300 nm and 500 nm simultaneously deposited on a flat GaAs substrate without fabricated Mie voids. Bottom, magnified view of the area highlighted by the black box. **b**, SEM image of the same area for comparison of the particle sizes. In the optical microscope image, particle size differences and interparticle distances can only be estimated approximately.

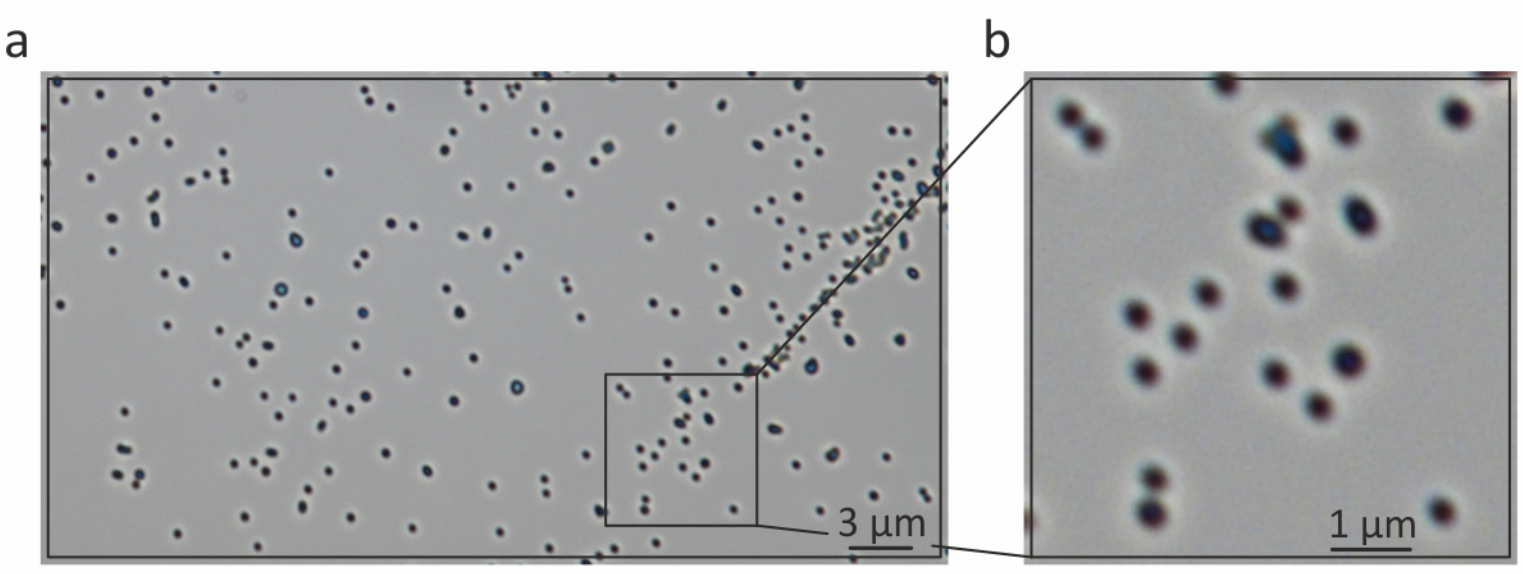

**SI 7: Comparison of PS and PMMA particles on a flat GaAs substrate.** **a**, Optical microscope image of PS and PMMA particles simultaneously deposited on a GaAs substrate without fabricated Mie voids. A clear distinction between the different particle types is not possible. **b**, Magnified view of the highlighted region supporting this observation.

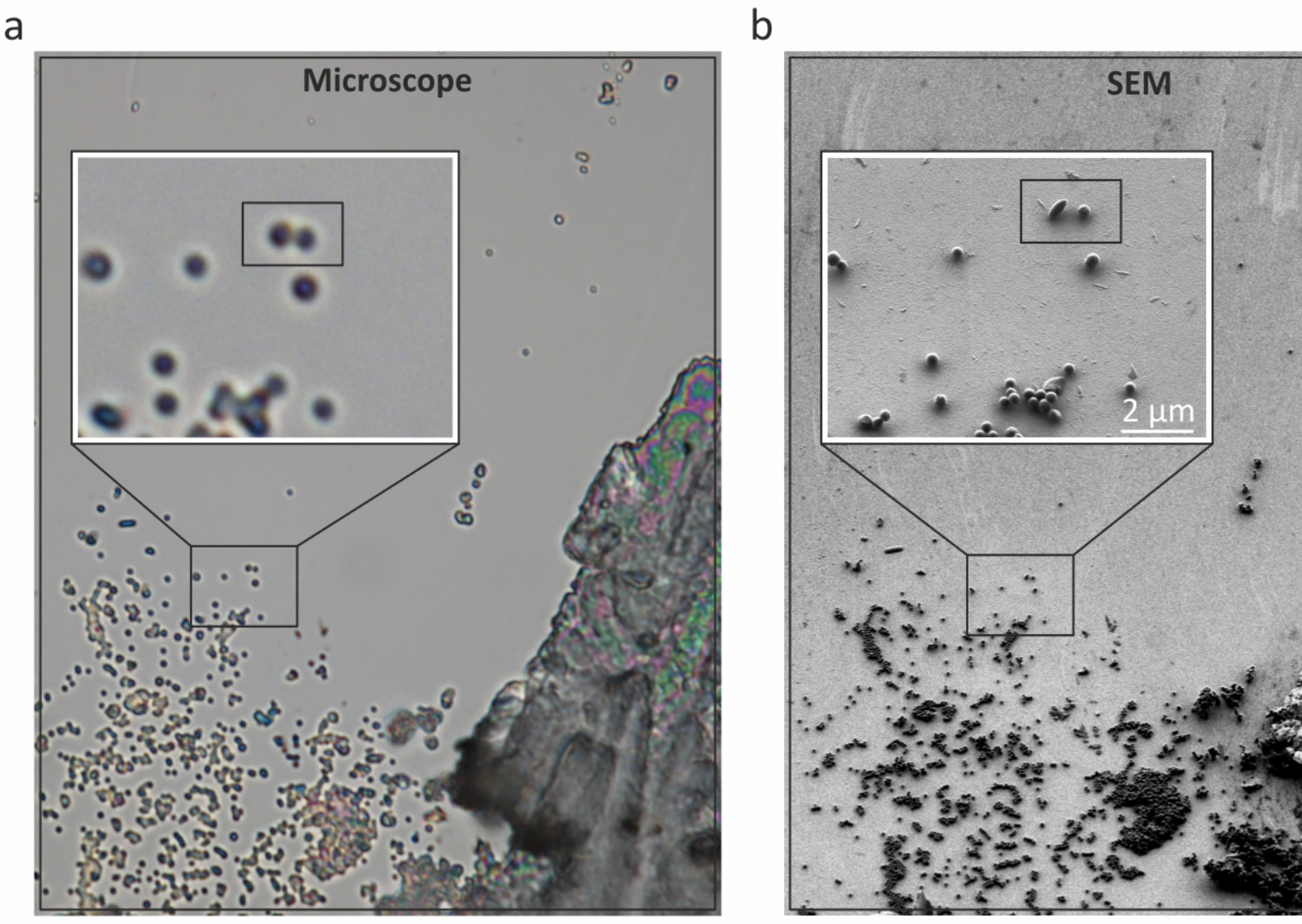

**SI 8: Investigation of spherical and elliptical particles on a flat GaAs substrate. a**, Optical microscope image of spherical and elliptical PS particles simultaneously deposited on a flat GaAs substrate without fabricated Mie void structures. The inset depicts a magnified view of the highlighted region, where no clear distinction between spherical and elliptical particles can be observed. **b**, SEM image of the same region confirming this observation. In the small inset, the black marked area contains a spherical and an elliptical particle in close proximity.

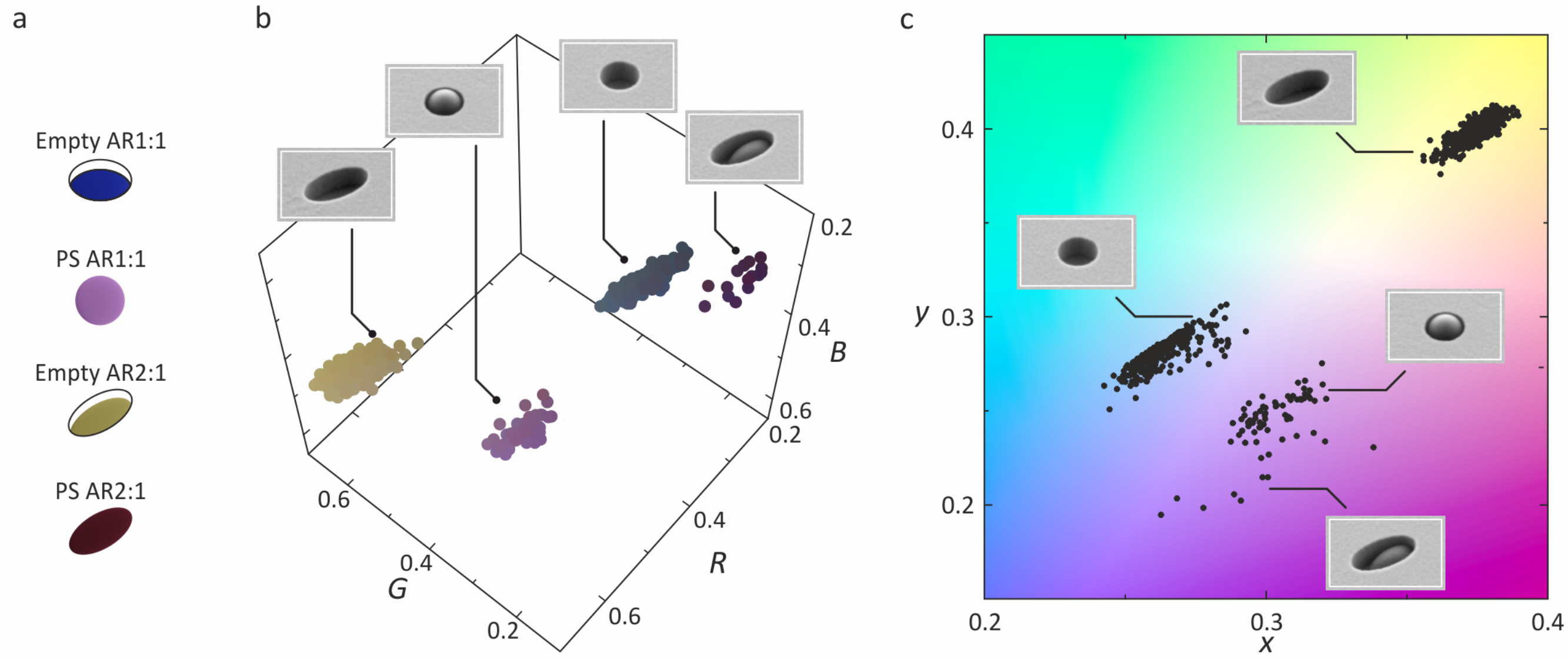

**SI 9: Color space comparison.** **a**, Four distinct void states from Figure 3 are displayed: empty and filled circular voids (blue and pink), and empty and filled elliptical voids with an aspect ratio (AR) of 2:1 (yellow and red), displayed using two different color representations. **b**, Three-dimensional RGB color space. **c**, CIE color diagram. The comparison of identical void colors across the two representations illustrates that an appropriate choice of color space enhances the separation and visibility of distinct color groups.

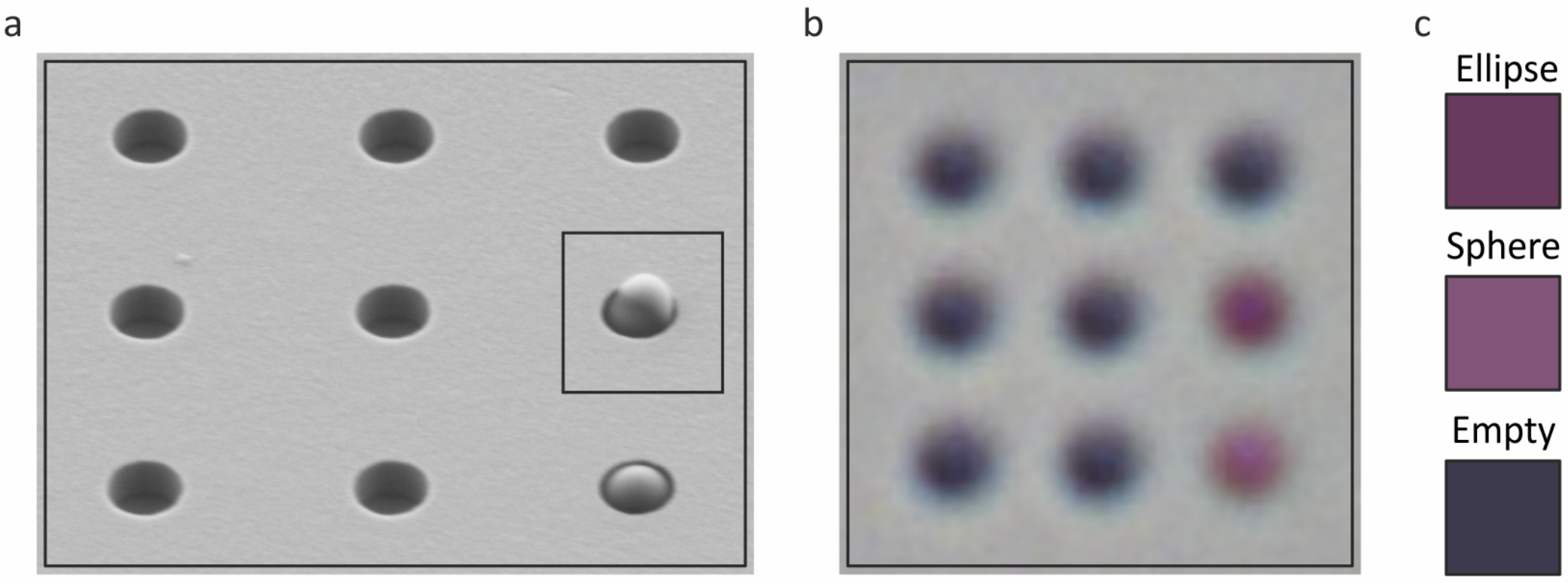

**SI 10: Investigation of elliptical particles in circular voids.** In rare cases, an elliptical particle can occupy a spherical void. **a**, SEM image of a 3 x 3 circular void array containing one spherical and one elliptical particle. The elliptical particle is indicated by the black square. **b**, Corresponding microscope image of the same region showing distinct colors for the two different particles. **c**, Representation of the colors corresponding to the void filled with an elliptical particle, with a spherical particle, and an empty void.

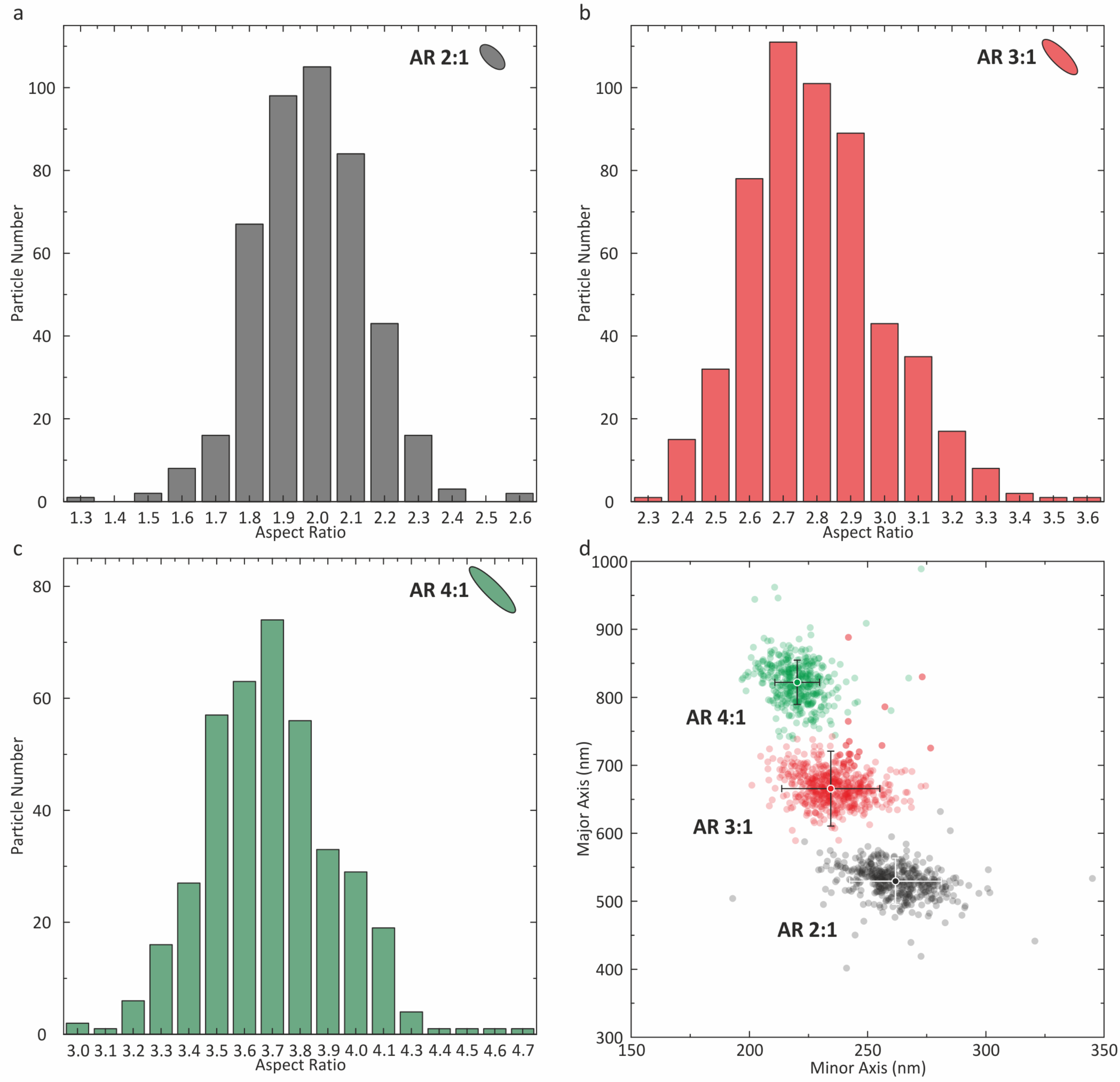

**SI 11: Detailed analysis of elliptical particles.** Histogram representation of the aspect ratios for particles designed with aspect ratios of **a** 2:1, **b** 3:1, and **c** 4:1. **d**, Absolute particle dimensions plotted as minor axis against the major axis for all particles (black: AR 2:1; red: AR 3:1; green: AR 4:1).

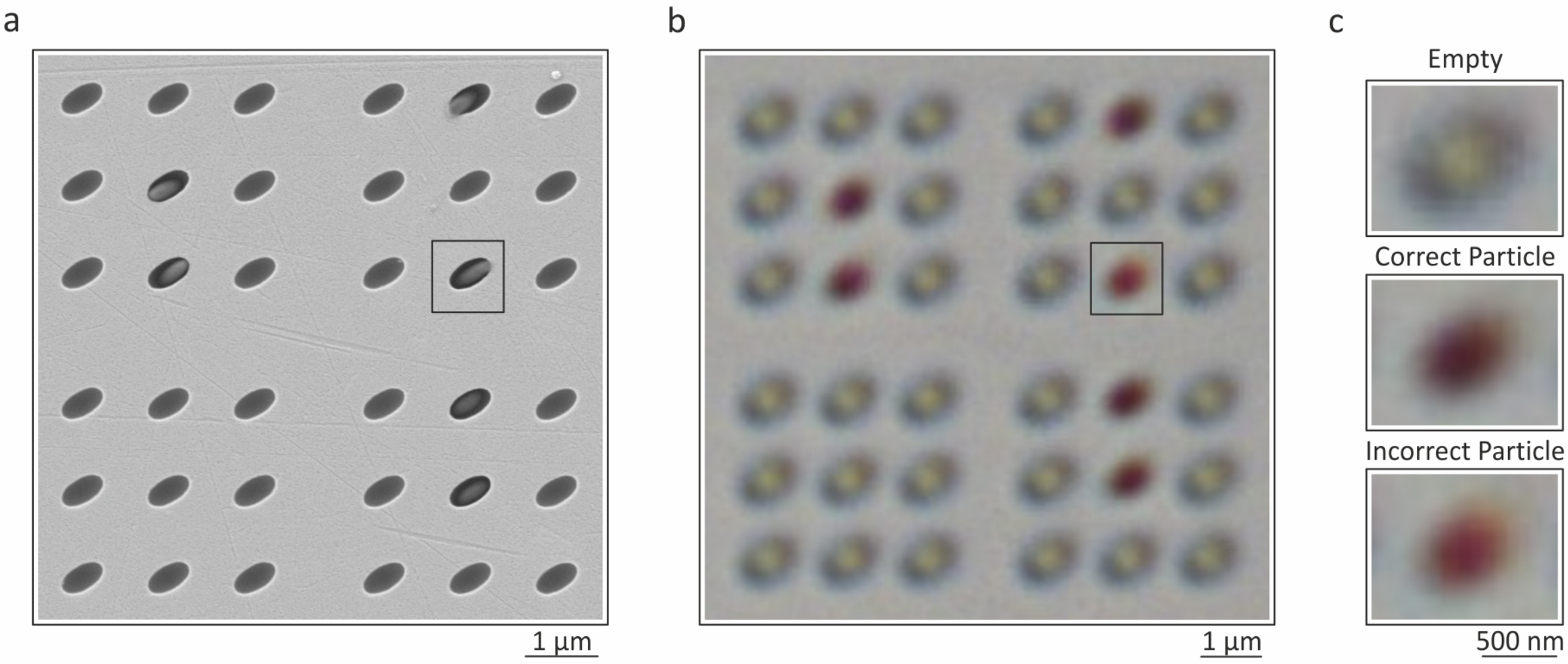

**SI 12: Investigation of elongated particles with incorrect aspect ratios in elliptical voids. a**, SEM image of four 3 x 3 void arrays. In the upper-right field, the indicated filled void contains an elliptical particle with an aspect ratio of 3:1, whereas all other elliptical particles exhibit an aspect ratio of 2:1. **b**, Corresponding microscope image of this area on the detection strip. The incorrectly filled void can be distinguished from the other filled voids. **c**, Magnified views on individual voids, showing an empty void, a correctly filled void, and the incorrectly filled void for comparison.

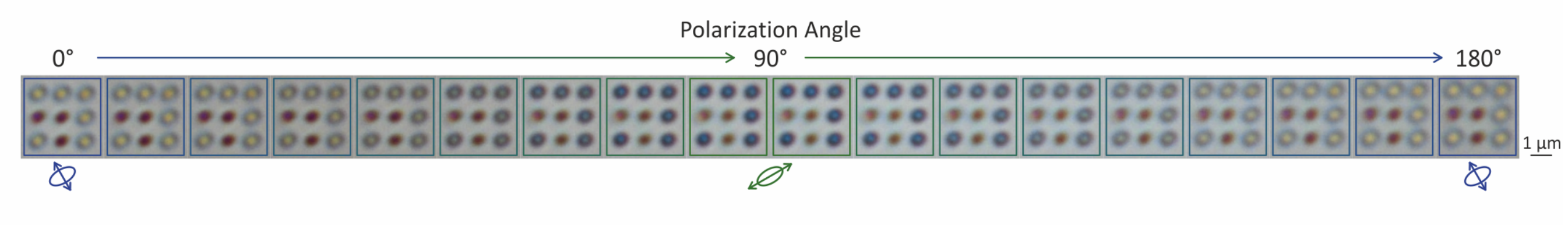

**SI 13: Presentation of all polarization states.** Optical microscope images used to determine the CIE distance presented in Fig. 5 are shown for all 18 different polarization states.

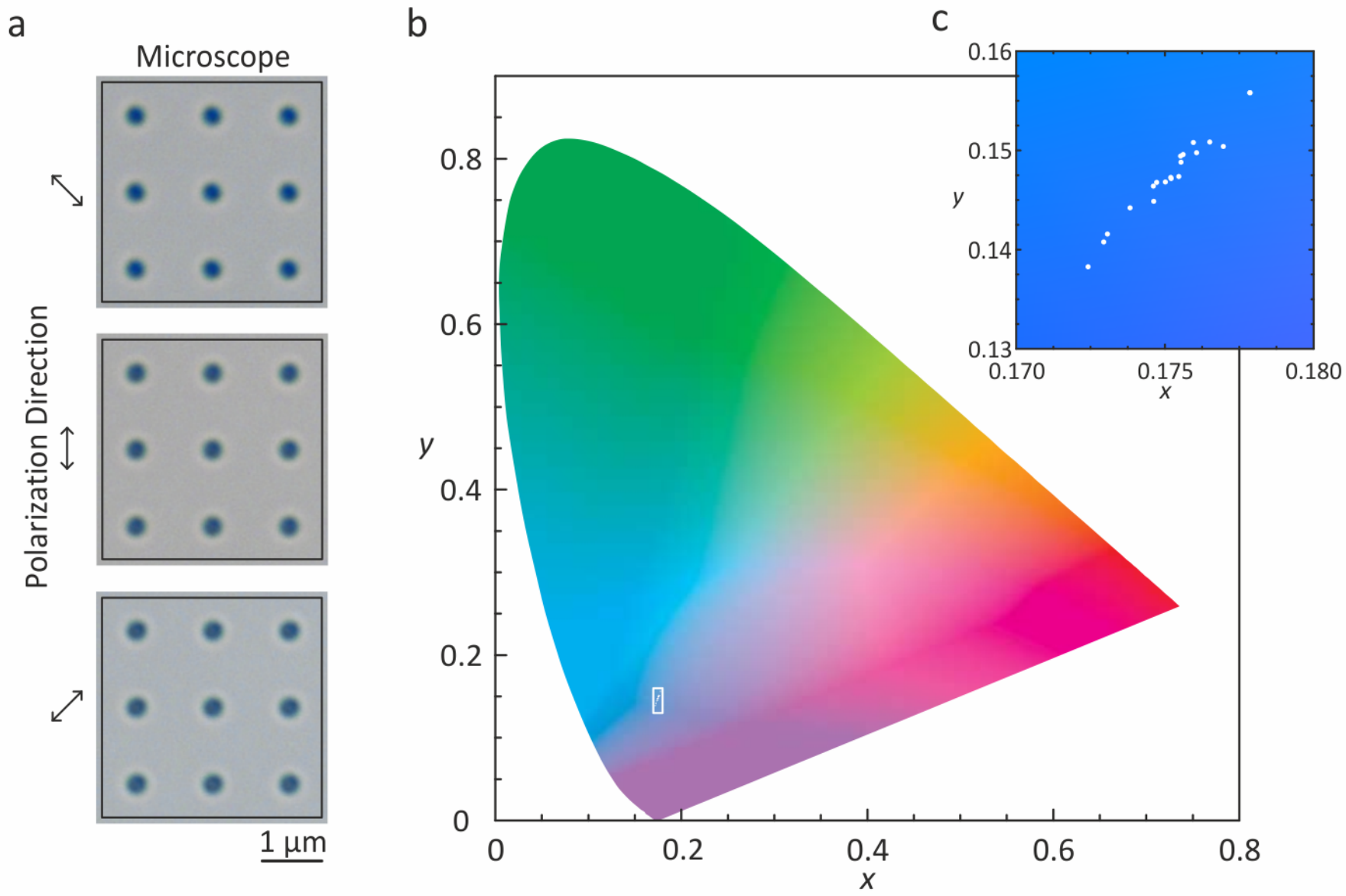

**SI 14: Polarization sensitivity of empty round voids.** **a**, Optical microscope images of nine empty round voids acquired for three different polarization states. **b**, Representation of the void colors for all polarization states in the CIE diagram, showing negligible color changes. **c**, Magnified view of the blue color range, highlighting the color stability of the round voids under polarization variation.

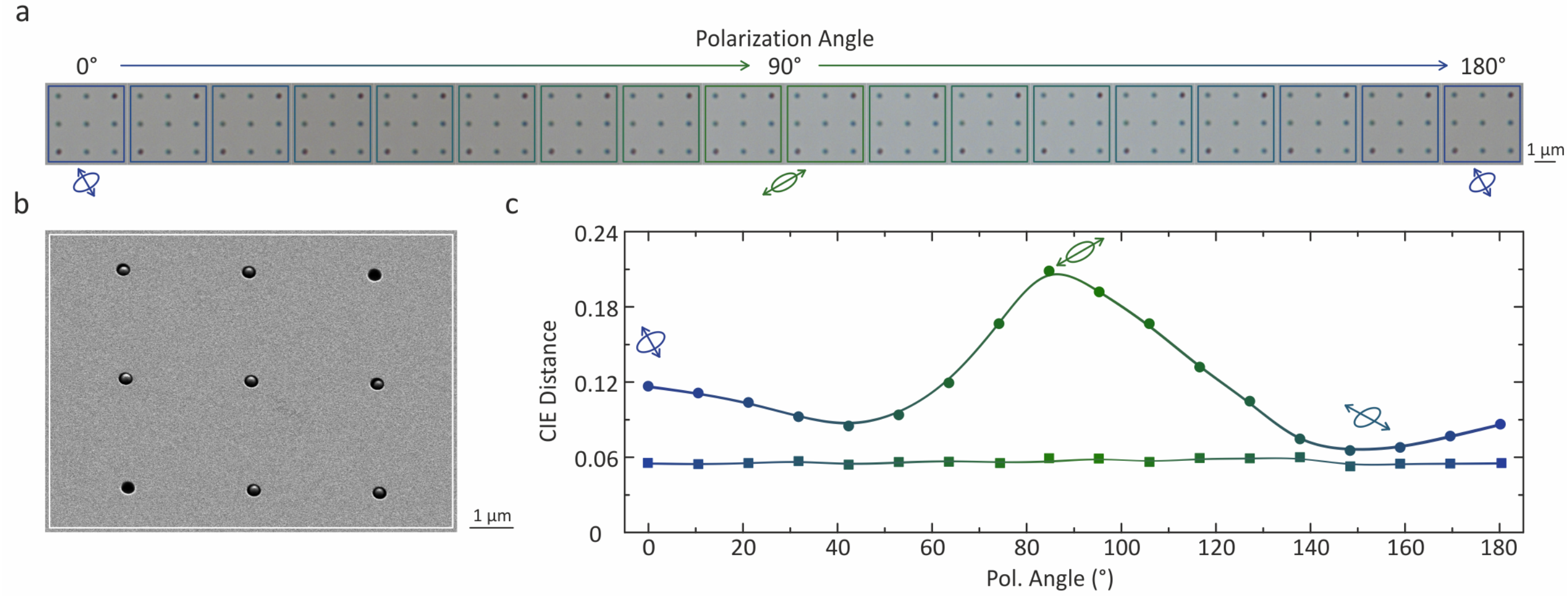

**SI 15: Polarization sensitivity of filled round voids.** **a**, Optical microscope images acquired for 18 different polarization states ranging from 0 ° to 180 °. **b**, SEM image of the corresponding field, indicating seven filled and two empty voids. **c**, CIE distance plotted as a function of the polarizer angle and compared with Figure 5g. Only negligible changes in the CIE distance are observed for the ideally filled round voids.

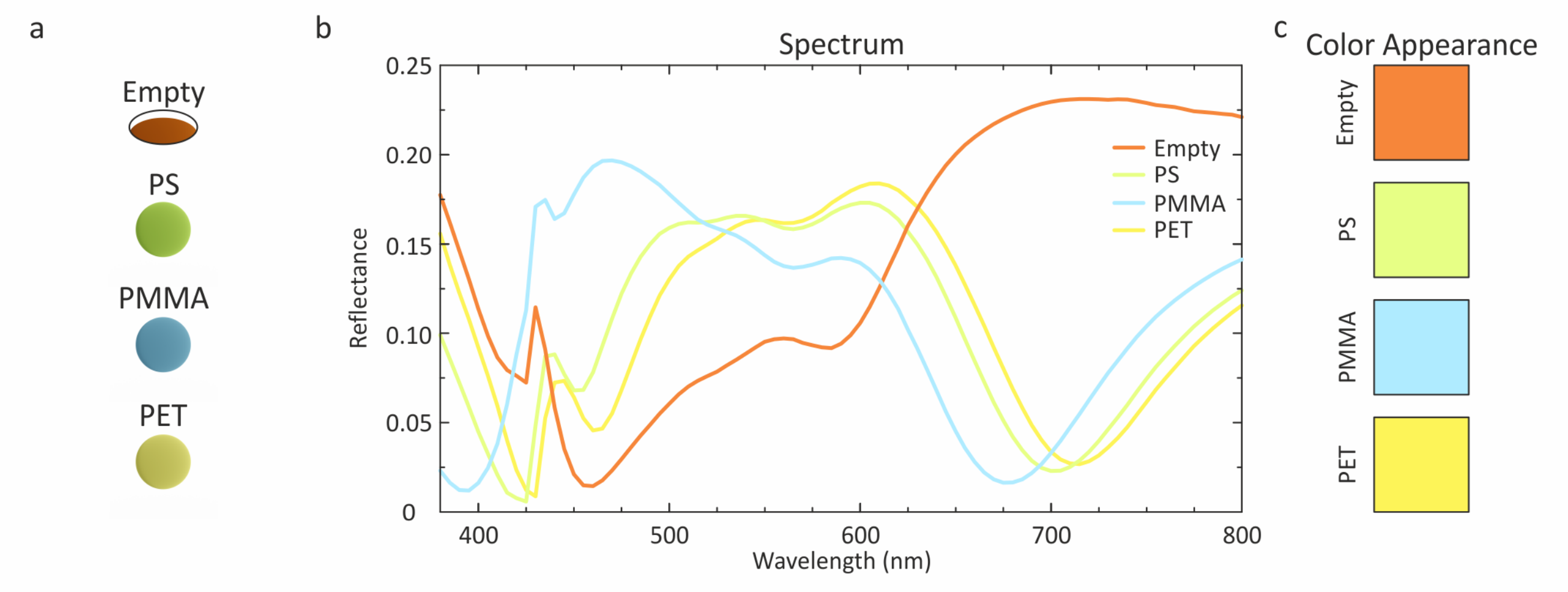

**SI 16: Simulated spectra and color appearances for different particle materials.** **a**, Simulations are performed for a 300 nm diameter empty void and voids filled with nanoparticles composed of PS, PMMA, and PET. **b**, Simulated reflection spectra, and **c**, Color appearance for an empty void and for a void filled with a spherical PS, PMMA, or PET particle. The void diameter is 300 nm with a depth of 200 nm.

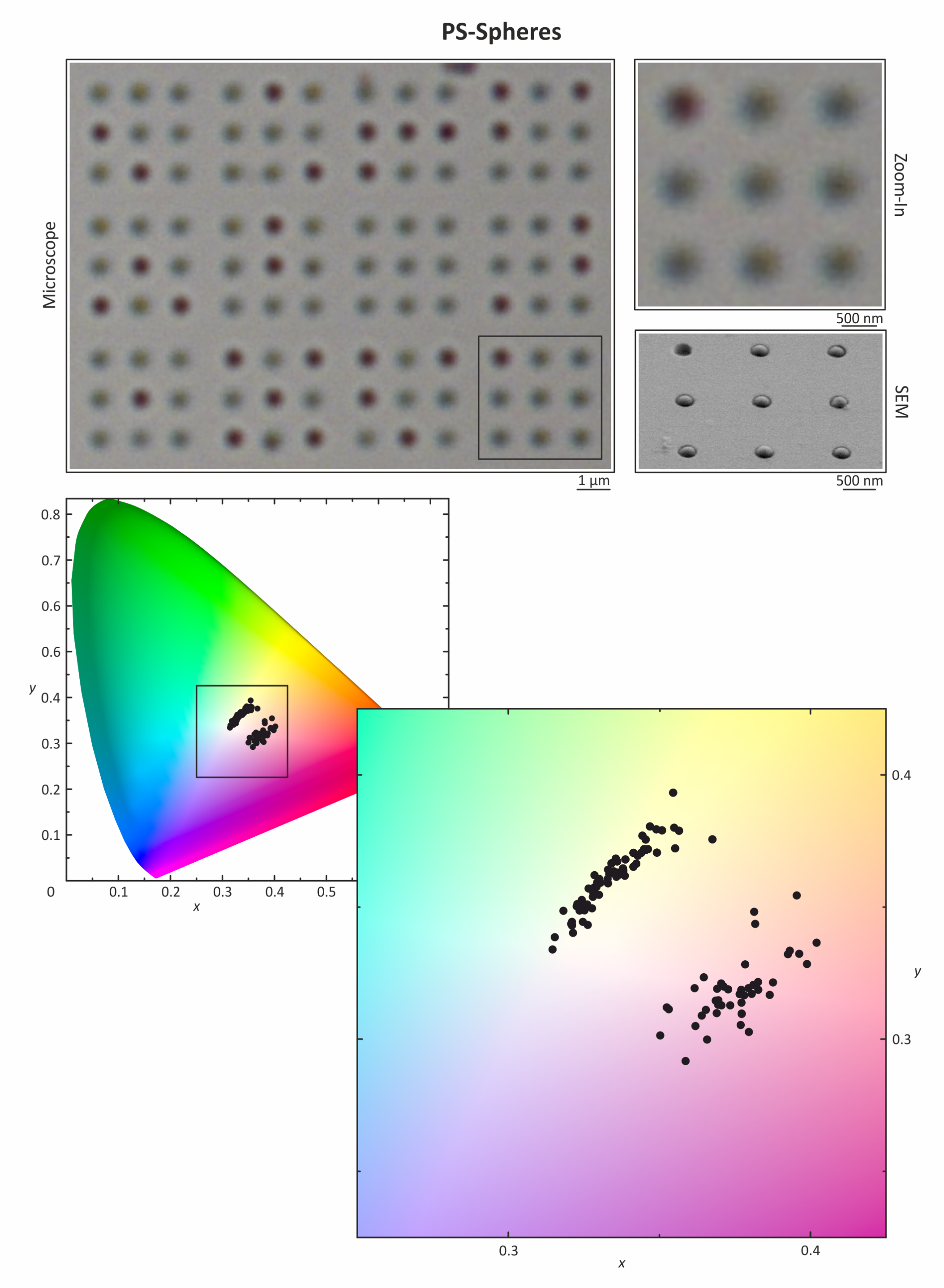

**SI 17: Second color reference measurement for PS particles.** PS particles are deposited to determine the characteristic color of voids filled with PS. (Top) Optical image of the detection strip with two magnified regions and corresponding SEM images. (Bottom) Full CIE-1931 chromaticity diagram and a magnified view of the relevant region, indicating a green color associated with PS-filled voids.

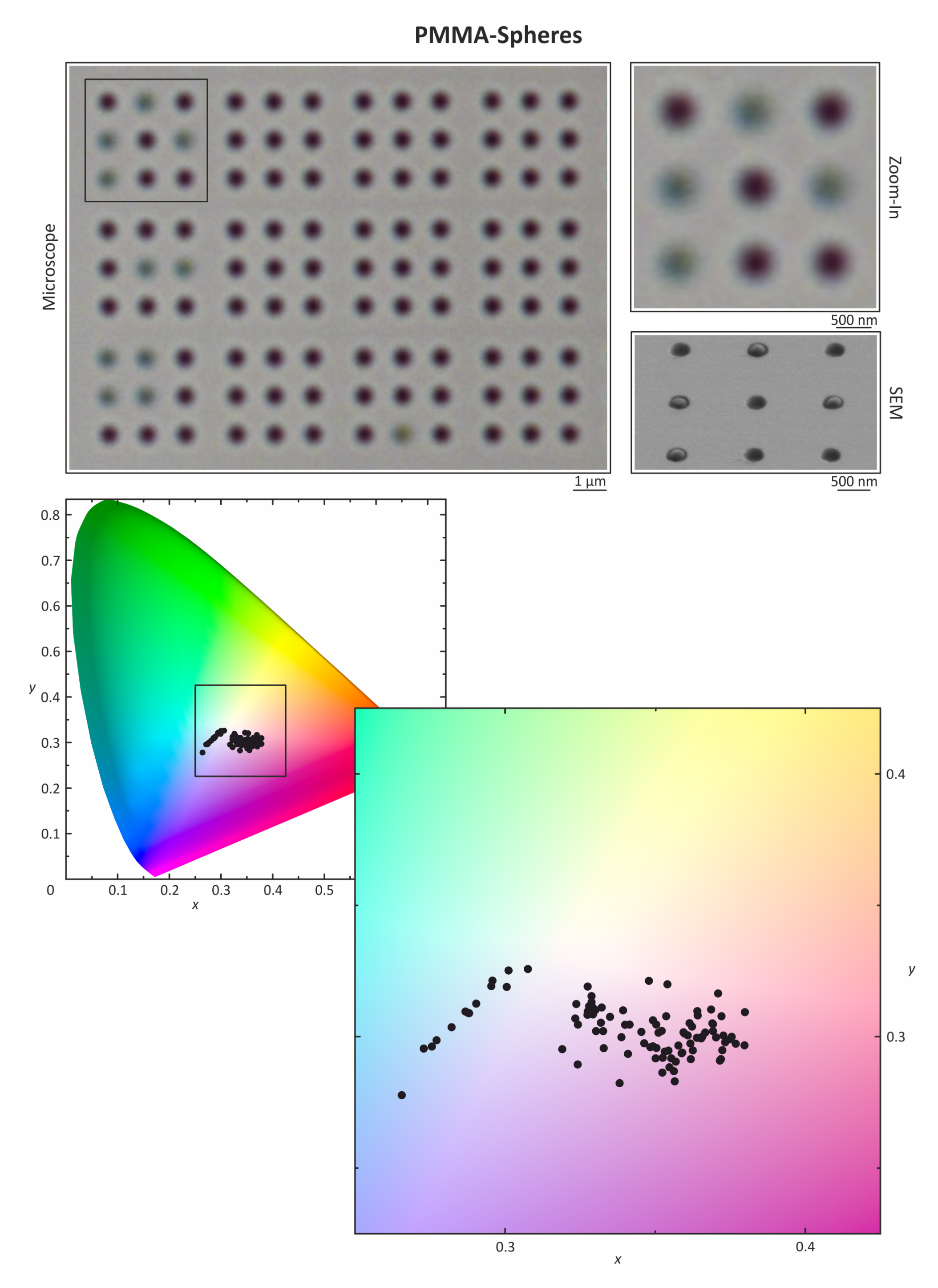

**SI 18: Second color reference measurement for PMMA particles.** PMMA particles are deposited to determine the characteristic color of voids filled with PMMA. (Top) Optical image of the detection strip with two magnified regions and corresponding SEM images. (Bottom) Full CIE-1931 chromaticity diagram and a magnified view of the relevant region, indicating a blue color associated with PMMA-filled voids.

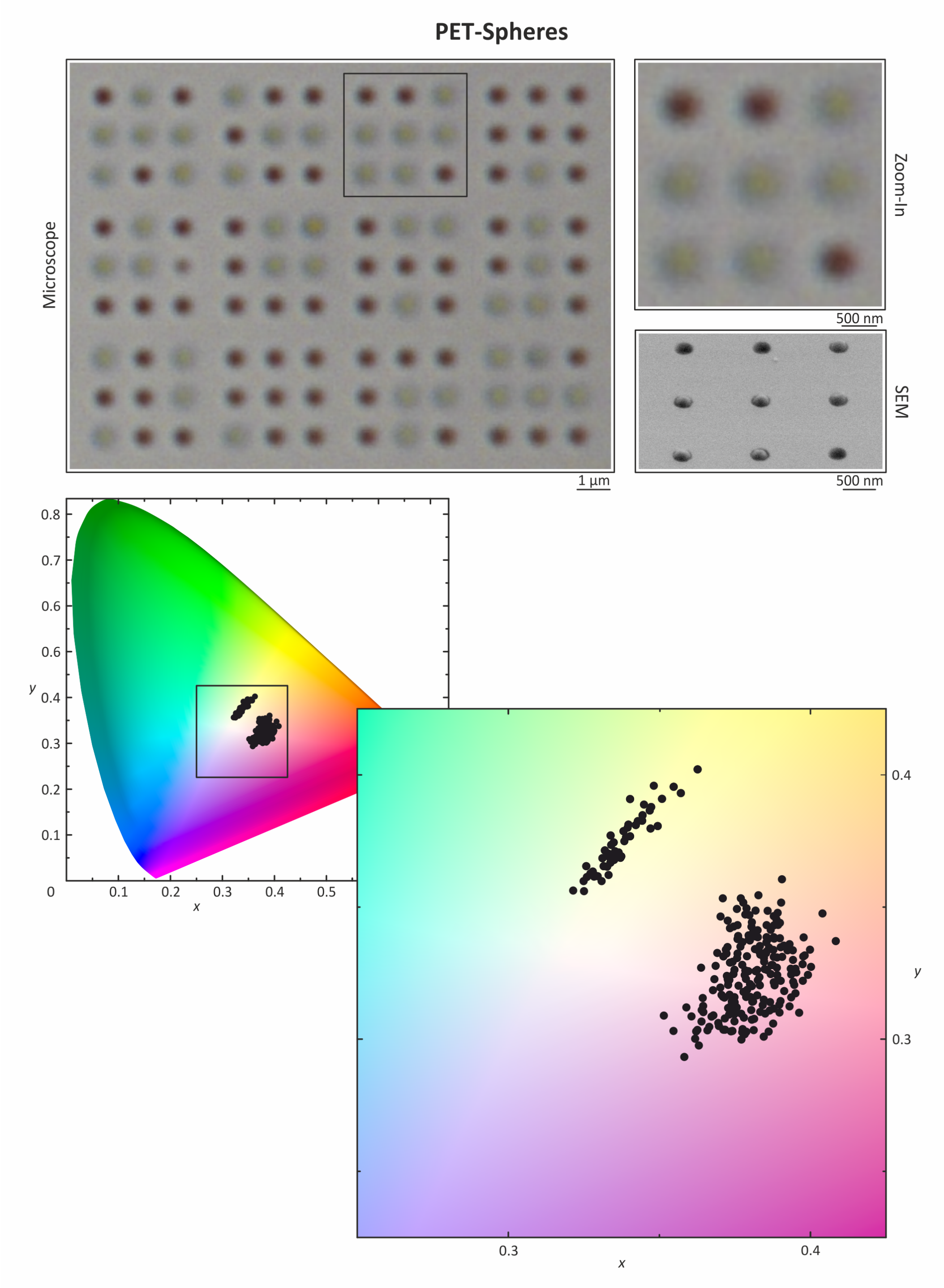

**SI 19: Second color reference measurement for PET particles.** PET particles are deposited to determine the characteristic color of voids filled with PET. (Top) Optical image of the detection strip with two magnified regions and corresponding SEM images. (Bottom) Full CIE-1931 chromaticity diagram and a magnified view of the relevant region, indicating a yellow color associated with PET-filled voids.